\documentclass[conference]{IEEEtran}

\usepackage{cite}
\usepackage{amsmath,amssymb,amsfonts}
\usepackage{algorithmic}
\usepackage{graphicx}
\usepackage{textcomp}
\usepackage{xcolor}
\def\BibTeX{{\rm B\kern-.05em{\sc i\kern-.025em b}\kern-.08em
    T\kern-.1667em\lower.7ex\hbox{E}\kern-.125emX}}

\usepackage{soul}
\usepackage{enumitem}
\usepackage[export]{adjustbox}
\usepackage[hyphens]{url}
\usepackage{lipsum}
\setlipsum{
  par-before = \begingroup\color{gray},
  par-after = \endgroup
}
\usepackage{ulem}
\usepackage{hyperref}
\hypersetup{colorlinks=true,linkcolor=blue,citecolor=blue,urlcolor=red}

\newcommand{\insertFigure}[5]{
    \begin{figure}[t]
      \centering
      \includegraphics[width=#3\linewidth]{figure/#1}
      \vspace{#4}
      \caption{\small #2}
      \label{fig:#1}
      \vspace{#5}
    \end{figure}
}

\newcommand{\insertFigureWide}[5]{
    \begin{figure*}[t]
      \centering
      \includegraphics[width=#3\linewidth]{figure/#1}
      \vspace{#4}
      \caption{\small #2}
      \label{fig:#1}
      \vspace{#5}
    \end{figure*}
}

\newcommand{\framework}{\textsc{Libra}}

\newcommand{\allreduce}{All-Reduce}
\newcommand{\alltoall}{All-to-All}
\newcommand{\reducescatter}{Reduce-Scatter}
\newcommand{\allgather}{All-Gather}

\newcommand{\equalBW}{EqualBW}
\newcommand{\perfBW}{PerfOptBW}
\newcommand{\perfCostBW}{PerfPerCostOptBW}

\newcommand{\nlg}{Turing-NLG}
\newcommand{\gpt}{GPT-3}
\newcommand{\msft}{MSFT-1T}
\newcommand{\resnet}{ResNet-50}
\newcommand{\dlrm}{DLRM}

\title{LIBRA: Enabling Workload-aware Multi-dimensional Network Topology Optimization for Distributed Training of Large AI Models}

\author{
    \IEEEauthorblockN{
        William Won\IEEEauthorrefmark{1},
        Saeed Rashidi\IEEEauthorrefmark{1}\IEEEauthorrefmark{4},
        Sudarshan Srinivasan\IEEEauthorrefmark{2},
        and Tushar Krishna\IEEEauthorrefmark{1}
    }
    \IEEEauthorblockA{
        \IEEEauthorrefmark{1}Georgia Institute of Technology, Atlanta, GA, USA\\
        \IEEEauthorrefmark{2}Intel, Bangalore, Karnataka, India
    }
    \IEEEauthorblockA{
        \IEEEauthorrefmark{1}william.won@gatech.edu 
        \IEEEauthorrefmark{1}saeed.rashidi@gatech.edu 
        \IEEEauthorrefmark{2}sudarshan.srinivasan@intel.com 
        \IEEEauthorrefmark{1}tushar@ece.gatech.edu
    }
}

\begin{document}
\maketitle

\begingroup\renewcommand\thefootnote{\textsection}
\footnotetext{Current affiliation: Hewlett Packard Enterprise [saeed.rashidi@hpe.com]}
\endgroup

\begin{abstract}

As model sizes in machine learning continue to scale, distributed training is necessary to accommodate model weights within each device and to reduce training time. However, this comes with the expense of increased communication overhead due to the exchange of gradients and activations, which become the critical bottleneck of the end-to-end training process. In this work, we motivate the design of multi-dimensional networks within machine learning systems as a cost-efficient mechanism to enhance overall network bandwidth. We also identify that optimal bandwidth allocation is pivotal for multi-dimensional networks to ensure efficient resource utilization. We introduce \framework, a framework specifically focused on optimizing multi-dimensional fabric architectures. Through case studies, we demonstrate the value of \framework, both in architecting optimized fabrics under diverse constraints and in enabling co-optimization opportunities.

\end{abstract}

\begin{IEEEkeywords}
multi-dimensional networks, distributed training, large language models, quadratic programming
\end{IEEEkeywords}

\section{Introduction}\label{sec:Introduction}

The demand for sophisticated, large-scale Machine Learning (ML) models is greater than ever. This is exemplified by recent Large Language Models (LLMs) powering tools like OpenAI ChatGPT~\cite{chatgpt}, Google Gemini~\cite{googlegemini}, and Microsoft Copilot~\cite{mscopilot}. The proliferation of ML-based applications has led to a growing trend of designing specialized High-performance Computing (HPC) systems to train ML models. Examples include Google Cloud TPU~\cite{googleCloudTpu}, Intel Habana HLS~\cite{intelHabana}, Meta Research SuperCluster~\cite{metarsc}, Tesla Dojo~\cite{TeslaDojo}, Cerebras CS-2~\cite{CerebrasCS2}, and Tensorrent Galaxy~\cite{tenstorrent}. Such Artificial Intelligence (AI) systems possess two notable features: (i)~Neural Processing Units (NPUs, e.g., GPUs or custom ASICs~\cite{tpuchip,HabanaPtP,TeslaD1,tenstorrent}) designed to efficiently execute compute operations, and (ii)~specialized network fabrics to scale-up and scale-out the system to thousands of NPUs. It is important to note that many of these AI systems are optimized at design-time for specific models of interest. For instance, the Meta ZionEX cluster~\cite{zionex} is specifically designed for accelerating large-scale DLRM~\cite{DLRM} training. While substantial research has been conducted on the architecture of NPUs~\cite{sze2017hardware,krishna2020data}, \textit{there has been limited work on architectural decisions for the custom fabric}, which constitutes the central focus of this work.

\insertFigure{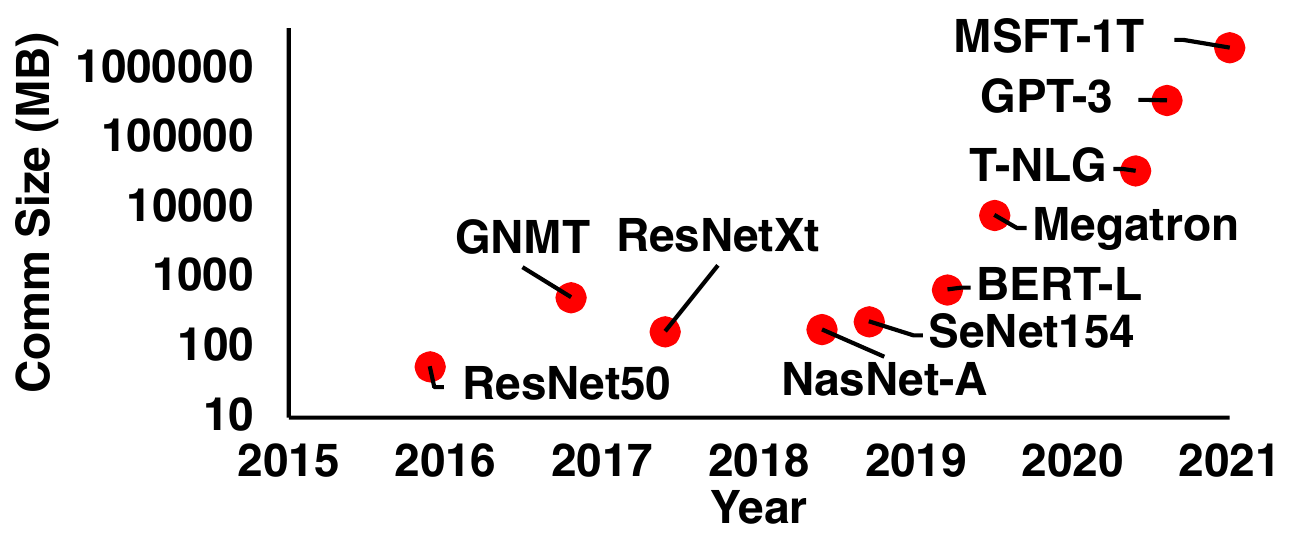}{
Communication sizes (FP16 as the datatype) for ML model training across 1,024 NPUs. The parallelization strategy of \nlg\ and smaller workloads are data parallel (minibatch size of 32), while \gpt\ and \msft\ use both tensor and data parallelism.
}{0.95}{-1.2em}{-0.5em}

\insertFigure{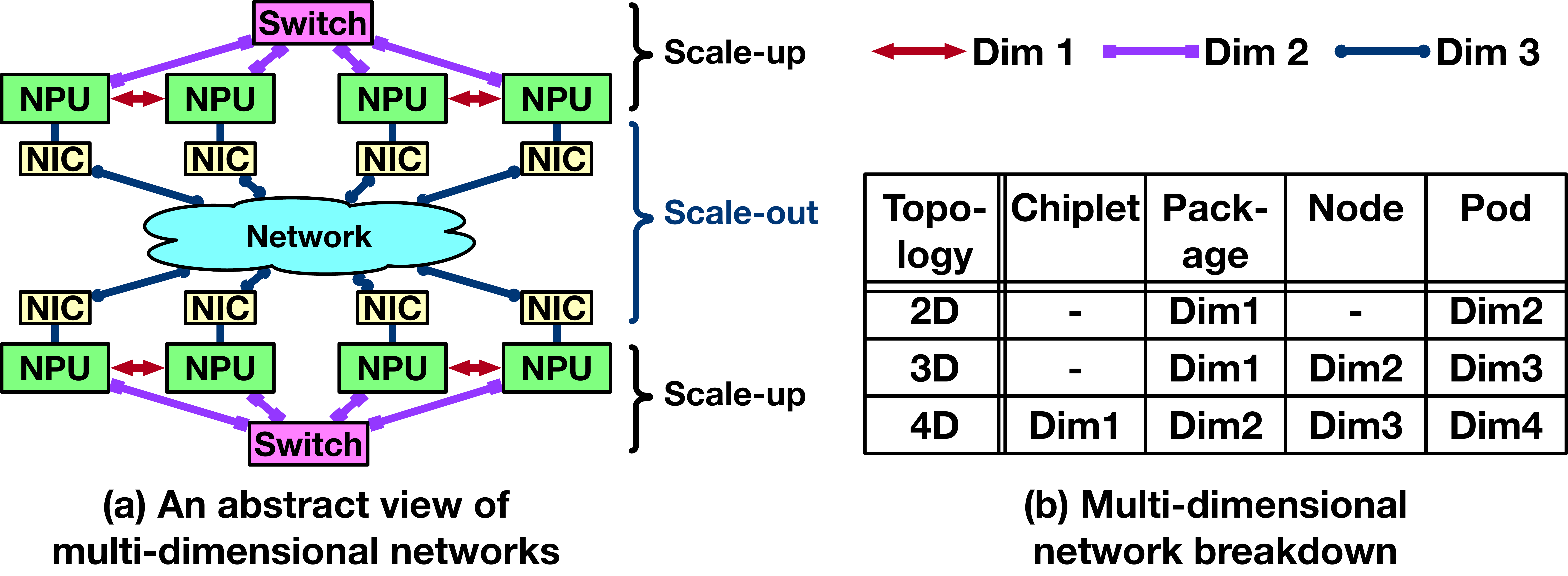}{
(a)~An abstract view of multi-dimensional networks. (b)~An example of physical connotation assigned to 2--4D networks.
}{1}{-2em}{-1.5em}

\insertFigureWide{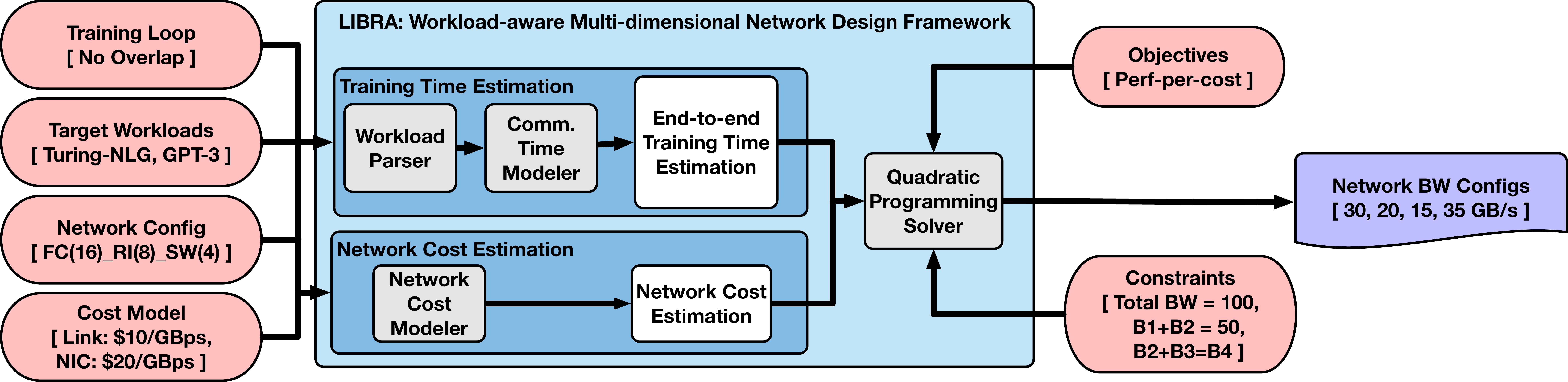}{
The architecture of \framework, a workload-aware multi-dimensional network BW optimization framework. Inputs to the \framework\ framework are represented in obrounds, with corresponding example values shown in square brackets. \framework\ estimates the end-to-end training time based on the provided network shape, training loop, and target workloads. Additionally, it calculates the cost of a network using a specified cost model and network shape. Subsequently, \framework\ searches for the optimal network BW configuration that maximizes the given objective while adhering to designated design constraints.
}{0.99}{-1em}{-1.7em}

AI models, especially LLMs, consist of billions to trillions of parameters (i.e., weights) that need to be distributed across NPUs since they cannot fit within the memory of a single device.\footnote{Even if a model fits within a single NPU, the raw compute requirement to train a model necessitates distributed training. For example, \gpt~\cite{gpt3} takes 355 years to be trained using a single NVIDIA V100 GPU~\cite{fbfsdp}.
} The synchronization of model activations and gradients during training leads to heavy communication among NPUs. As shown in \autoref{fig:CommSizes}, the total communication size for large model training can span GBs to TBs. Consequently, communication becomes one of the major bottlenecks in distributed training~\cite{dlCollective,CommBottleneck1,CommBottleneck2,CommBottleneck3, saeedACE}. This naturally necessitates driving higher network bandwidth (BW) resources per NPU to efficiently process the massive volume of communication during training.

Therefore, AI clusters in data centers today most commonly leverage two-dimensional (i.e., 2D) network fabrics to achieve high network BW. In the first dimension (i.e., Dim), they employ custom rack-to-rack links (e.g., XeLink~\cite{xelink} or NVLink~\cite{NVLink3}) to scale-up a node, enabling peer-to-peer communication among NPUs. The second dimension is the conventional scale-out fabric, which employs Network Interface Cards (NICs) using technologies such as Ethernet~\cite{NIC400G} or InfiniBand~\cite{nvidiainfiniband}. However, as AI models continue to scale at an unprecedented rate~\cite{kaplan2020scaling}, next-generation AI systems demand even higher network BW, and addressing this challenge remains an open problem. A naive approach would be to increase the individual BW of the two dimensions. However, it is pivotal to note that physical and management constraints, such as limited pin counts allotted per each network dimension, electrical SerDes constraints, signaling, and power consumption, cap the potential network BW achievable per each network dimension~\cite{maximumNPUBW}. For example, while we can expect some further increase in BW across generations of scale-up and scale-out links, NVLink technology today only reaches a maximum of 450 GB/s~\cite{NVLink3}. This is true even with transitioning to alternative technologies that may offer higher BW, such as advanced wafer-scale packaging~\cite{mcm-gpu,simba,waferScaleGPU} or photonics~\cite{tpuv4,passage}, since additional physical constraints like thermals and reliability still apply as well as expensive manufacturing costs (at least today). In summary, there is no single technology that can provide high network BW within a single network dimension. This is evident from the diverse fabrics used in AI systems today: Intel leverages XeLink and InfiniBand technologies~\cite{intelsystem}, NVIDIA uses NVLink, NVSwitch, and InfiniBand NICs~\cite{dgx2, A100, NvidiaH100}, and Cerebras~\cite{CerebrasCS2} and Tesla~\cite{TeslaDojo} utilize Wafer-scale and Chiplets.

We believe that a promising approach to enhance the BW per NPU is to (i)~explicitly add more network dimensions, and (ii)~leverage a mixture of fabric technologies. An abstract view of such multi-dimensional fabrics for AI systems is demonstrated in \autoref{fig:HierarchicalTopologyDefinition}. By having multiple network dimensions, we can overcome the constraint of limited BW of each network dimension and drive overall higher network BW per NPU. For example, Google Cloud TPUv4~\cite{tpuv4} employs a 3D point-to-point electrical network augmented by a photonic network dimension in order to drive more network resources~\cite{tpuv4}. Even when employing a multi-dimensional network to allocate equal BW resources per NPU, it can still achieve a higher performance-per-cost (i.e., perf-per-cost) design point. This is attributable to the load-reducing nature of the multi-rail communication algorithm, which enables substituting expensive network resources with more cost-efficient technologies.

This multi-dimensional network scheme opens up a new fabric architecture optimization problem that this work aims to solve: \textit{how to determine BW distribution across different dimensions at design-time, under diverse technology-driven dollar cost and BW constraints, while enabling high performance across multiple AI workloads?} As we show in this work, judicious optimization of network BW at design-time is necessary to enable co-optimization and scheduling opportunities at runtime. Otherwise, BW resources augmented by leveraging multi-dimensional topologies may be underutilized, resulting in inefficiencies in communication and training slowdown~\cite{themis}. Resource allocation design is always a challenging problem, akin to other design challenges like sizing scratchpads or caches. However, unlike general-purpose HPC applications, communication patterns (e.g., communication types, sizes, and the group of NPUs involved) for ML applications can be predetermined when the parallelization strategy of a workload is set. Therefore, focusing on AI workloads enables us to adopt a more systematic workload-aware approach.

To this end, we propose \textbf{\framework: \uline{L}everaging \uline{I}ntelligent \uline{B}andwidth \uline{R}esource \uline{A}llocation for multi-dimensional networks}.\footnote{
\textit{Libra} also symbolizes \textit{balance} in Latin, aligning with this work of judiciously allocating and harmonizing BW resources across dimensions.
} \framework\ is a workload-aware, multi-dimensional network optimization framework at design-time.\footnote{
\framework\ yields an optimized network \textit{design} targeted for a specific set of workloads. We use the term \textit{design-time} to contrast \framework, as a toolchain, with runtime-based methodologies (e.g., schedulers or load balancers).
} The architecture of \framework\ is summarized in \autoref{fig:LibraFramework}. Given a set of target Deep Neural Network (DNN) models, fabric technology options, and design constraints, \framework\ can swiftly estimate the training performance and propose the optimal network BW design point that maximizes performance or perf-per-cost. Through case studies, we demonstrate that \framework\ can be used for two primary purposes: (i)~optimizing the multi-dimensional network architecture for a family of target workloads, and (ii)~exploring co-optimization opportunities, such as designing networks alongside target workload parallelizations or runtime-based training scheduling techniques. To the best of our knowledge, \framework\ is the first workload-aware constrained-optimization framework for AI systems fabric design.

\section{Background}\label{sec:background}

\insertFigure{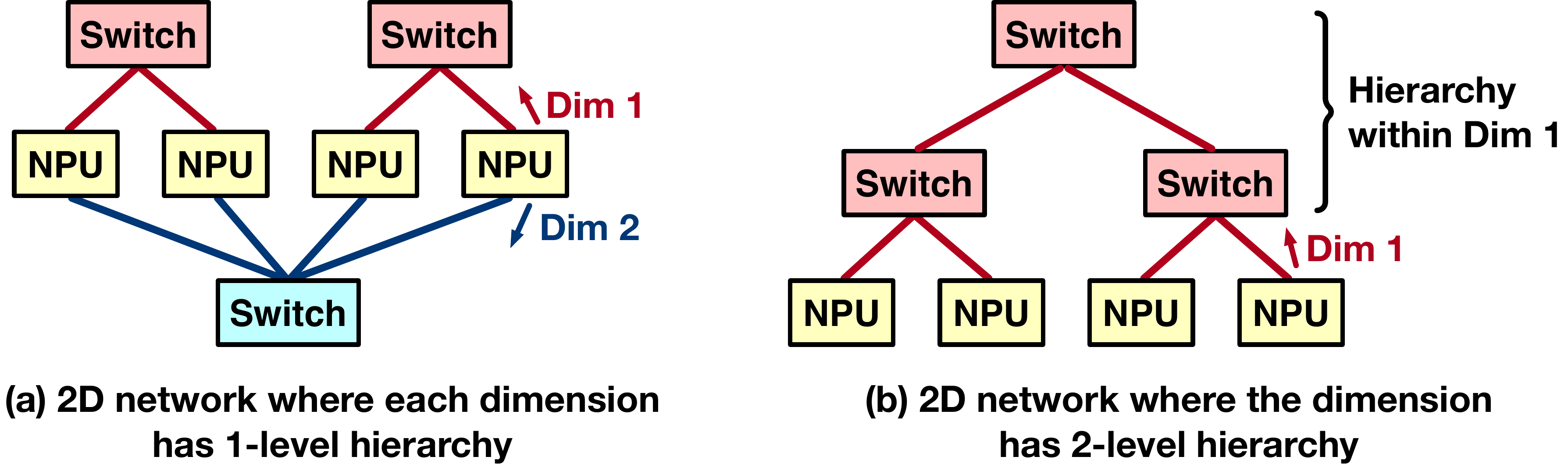}{
(a)~An example of a switch-based 2D network. (b)~Example 1D topology, which utilizes a 2-level hierarchical switch in Dim 1.
}{1}{-2em}{-1.5em}

\subsection{Multi-dimensional Network}

A \textit{multi-dimensional network} is defined as a network topology where each NPU has multiple independent connectivity options, which can be accessed in parallel to communicate with other NPUs. An abstract view of the multi-dimensional network is illustrated in \autoref{fig:HierarchicalTopologyDefinition}(a). This concept is equivalent to the definition of multi-rail networks~\cite{multirail1, multirail2}. In \autoref{fig:HierarchicalTopologyDefinition}(b), physical technology connotations are assigned to each network dimension as an illustrative example for 2--4D networks.\footnote{
This example physical connotation is used for evaluation purposes in this paper. Still, \framework\ remains flexible and supports arbitrary dimensionalities.} 
We introduce Chiplet, Package, Node, and Pod, inspired by emerging technology and system trends. Chiplet represents an NPU chip. Package consists of one or more Chiplets interconnected via Multi-chip Module (MCM) packaging~\cite{simba, IntelGPU}. A Pod is an inter-server scale-out unit used in some platforms today, typically interconnected through NICs~\cite{intelHabana}. A Node constructs a server unit by connecting multiple Packages through an inter-board network.

It's worth noting that in our terminology, a network \textit{dimension} is distinct from adding a \textit{hierarchy} within a network dimension. Each dimension of the network is directly accessible to the NPU (via explicit ports/pins) in parallel. In contrast, within a dimension, the implementation may choose to use a hierarchy, such as a hierarchy of switches instead of a single large switch. This distinction is depicted in \autoref{fig:DimensionHierarchyComparison}. While both topologies utilize three physical switches, \autoref{fig:DimensionHierarchyComparison}(a) represents a 2D network since NPUs have two distinct switch-based networks accessible in parallel, whereas \autoref{fig:DimensionHierarchyComparison}(b) showcases a 1D topology using a 2-level switch hierarchy within Dim 1.

\subsection{Distributed Training}

\noindent \textbf{Parallelization Strategy.} Modern foundation models, along with other large models, often have a substantial memory footprint during training, which exceeds the capacity of a single NPU memory~\cite{mszero}. To address this issue, the model and training dataset need to be sharded and distributed across multiple NPUs. The \textit{parallelization strategy} governs how the model and dataset are divided and placed. There are two main parallelization strategies: Tensor Parallelism (TP) and Data Parallelism (DP). TP divides the model and distributes them across NPUs, reducing the memory requirement of each NPU during training~\cite{surveyml}. We use TP-$n$ to denote that the model is sharded in $n$-way. DP distributes the training dataset to boost the training throughput~\cite{surveyml}. We use DP-$n$ to indicate that the training dataset is split into $n$ separate groups. TP and DP are orthogonal and can be combined, resulting in a hybrid parallelism (HP) scheme. We use HP-$(m, n)$ to denote the mixture of TP-$m$ and DP-$n$. In the HP-$(m, n)$ setup, the training dataset is first split into $n$ sets, and each set is fed into the group where the model is $m$-way sharded. Consequently, HP-$(m, n)$ requires a total of $m \times n$ NPUs. We leverage sophisticated hybrid parallelization schemes such as Megatron-LM~\cite{megatronlm}, combined with the ZeRO-2 optimizer~\cite{mszero}.

\insertFigure{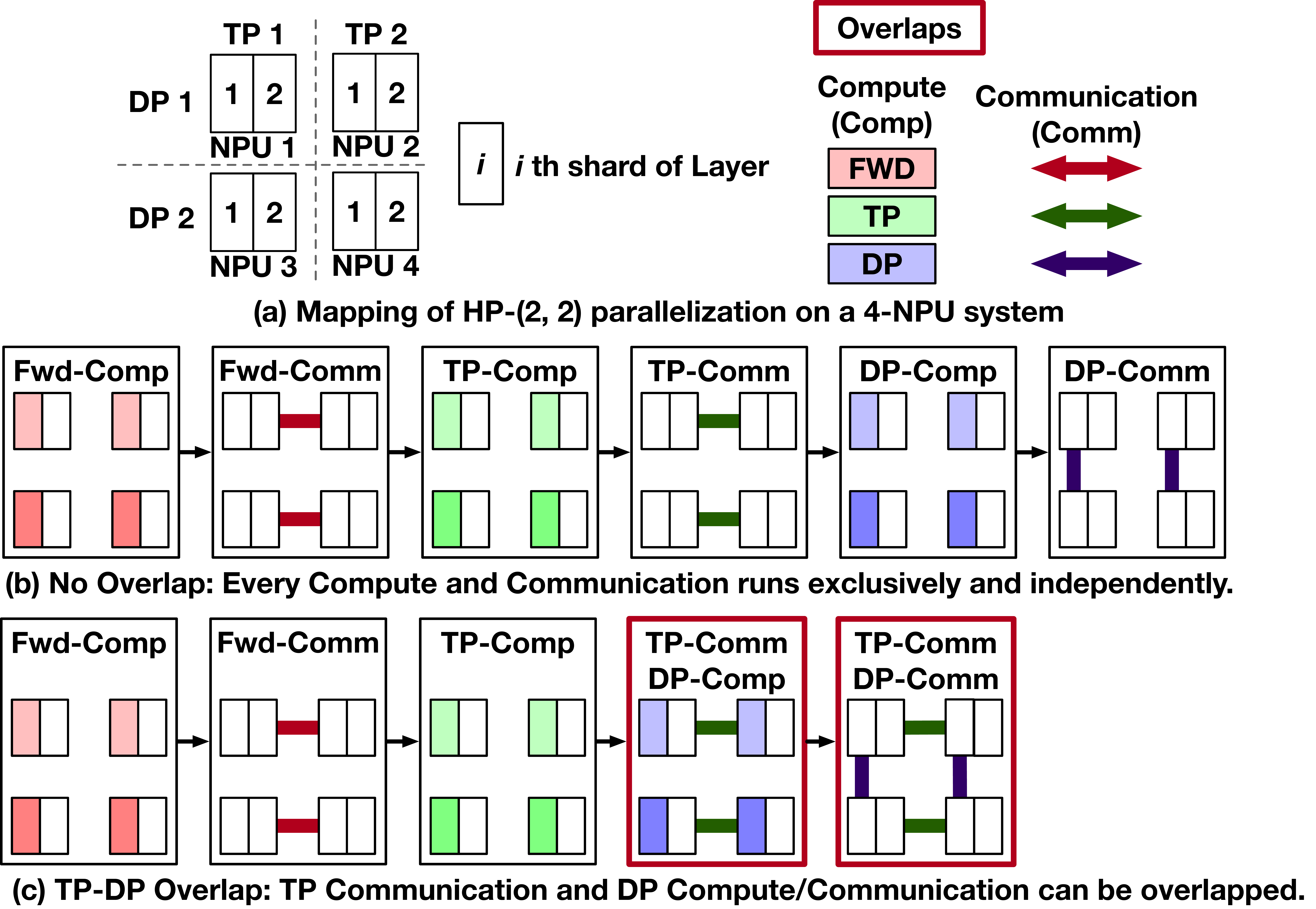}{
Examples of training loops. (a)~HP-(2,2) parallelization strategy (b)~A training loop without any overlap (No Overlap) (c)~A training loop with TP and DP running concurrently (TP-DP Overlap).
}{1}{-2em}{-0.5em}

\insertFigure{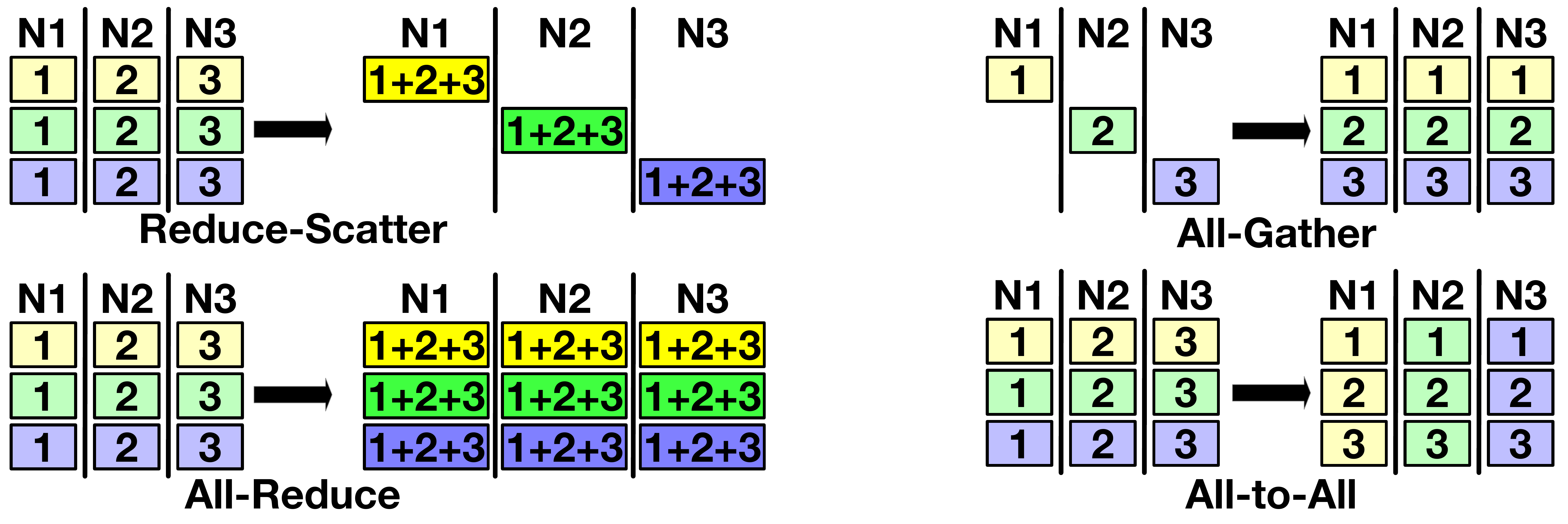}{
Common collective patterns (N denotes an NPU).
}{1}{-2.2em}{-1.5em}

\noindent \textbf{Training Loop.} Each parallelization strategy necessitates devices to communicate and synchronize dispersed information. For instance, TP requires input activations and input gradients to be communicated, while DP necessitates the synchronization of weight gradients. The \textit{training loop} defines the ordering and scheduling of computation and communication during the training process. As shown in \autoref{fig:TrainingLoop}, we provide two examples of training loops for HP-$(2, 2)$ of a single-layer model. In \autoref{fig:TrainingLoop}(b), there are no overlaps between the computation and communication stages. In contrast, \autoref{fig:TrainingLoop}(c) allows the overlap of TP communication and DP computation and communication, resulting in a shorter training time.

\subsection{Collective Communication}\label{subsec:collectivecomms}

\noindent \textbf{Topology-aware Collective Algorithms.} Communications required by parallelization strategies are managed through collective communications (i.e., collectives). Common collectives in distributed training are illustrated in \autoref{fig:Collective2x2}. The most prominent collective in distributed training is \allreduce~\cite{dlCollective}. It is logically equivalent to \reducescatter\ followed by an \allgather~\cite{astrasim}. In certain cases of TP, like embedding tables~\cite{DLRM}, \alltoall\ is required. Real systems execute collectives using collective communication algorithms through Collective Communication Libraries (CCLs)~\cite{nccl, oneccl}. For example, Ring, Direct, and Recursive Halving-Doubling are commonly used \allreduce\ algorithms~\cite{collective1}. These algorithms are \textit{topology-aware collective algorithms} designed for Ring, FullyConnected, and Switch networks, respectively, ensuring that they do not introduce link contention when running on their respective physical topologies. \autoref{fig:NetworkBuildingBlocks} lists common network building blocks and their corresponding topology-aware collective algorithms.

\noindent \textbf{Multi-rail Collective Algorithm}. Basic collective algorithms are not ideally suited for direct use over multi-dimensional networks. For instance, the Direct collective algorithm performs well on a FullyConnected network, but the physical connectivity of multi-dimensional networks often does not meet such expectations. Consequently, heavy network contention and oversubscription over low-BW links can occur, leading to significant underutilization of network BW. To address this issue, \textit{multi-rail} collective algorithms have been proposed to fully leverage the resources of multi-dimensional networks~\cite{blueconnect, astrasim}, which is the approach adopted in \framework. A multi-rail collective algorithm capitalizes on the inherent nature of multi-dimensional networks, where basic network building blocks are stacked up. Therefore, it executes basic collective algorithms in sequence. For example, to perform an \allreduce\ collective on an $N$-dimensional network:
\begin{itemize}[leftmargin=*]
    \item Run \reducescatter\ on Dim 1. Then, run \reducescatter\ on Dim 2. These \reducescatter\ jobs, in ascending order, continue up to Dim $N$ ($N$ \reducescatter\ stages).
    \item Perform \allgather\ on Dim $N$, then execute \allgather\ on Dim $N-1$. This \allgather\ stage continues, in descending order, down to Dim 1 ($N$ \allgather\ stages).
\end{itemize}

In summary, the multi-rail \allreduce\ collective algorithm involves a total of $2N$ stages. Within each stage, each dimension utilizes its corresponding topology-aware collective algorithm. This approach guarantees that the overall collective operation runs in a contention-free manner. An example \allreduce\ on a $3 \times 2$ (2D) network is illustrated in \autoref{fig:MultiDimCollExample}.

\insertFigure{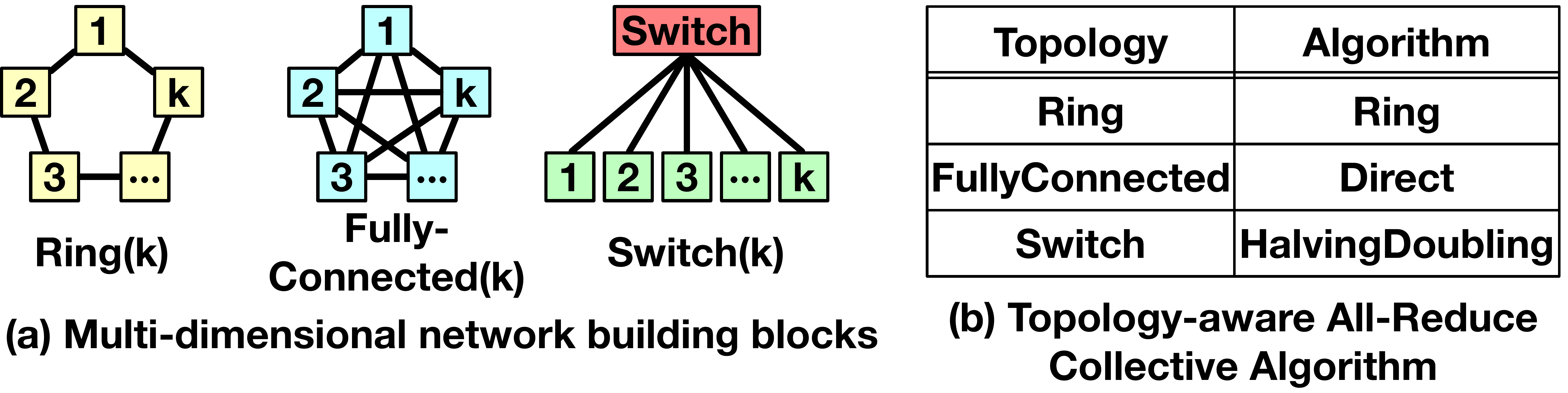}{
(a)~Multi-dimensional network building blocks. (b)~their corresponding topology-aware \allreduce\ algorithm.
}{1}{-2.3em}{-1.5em}

\insertFigure{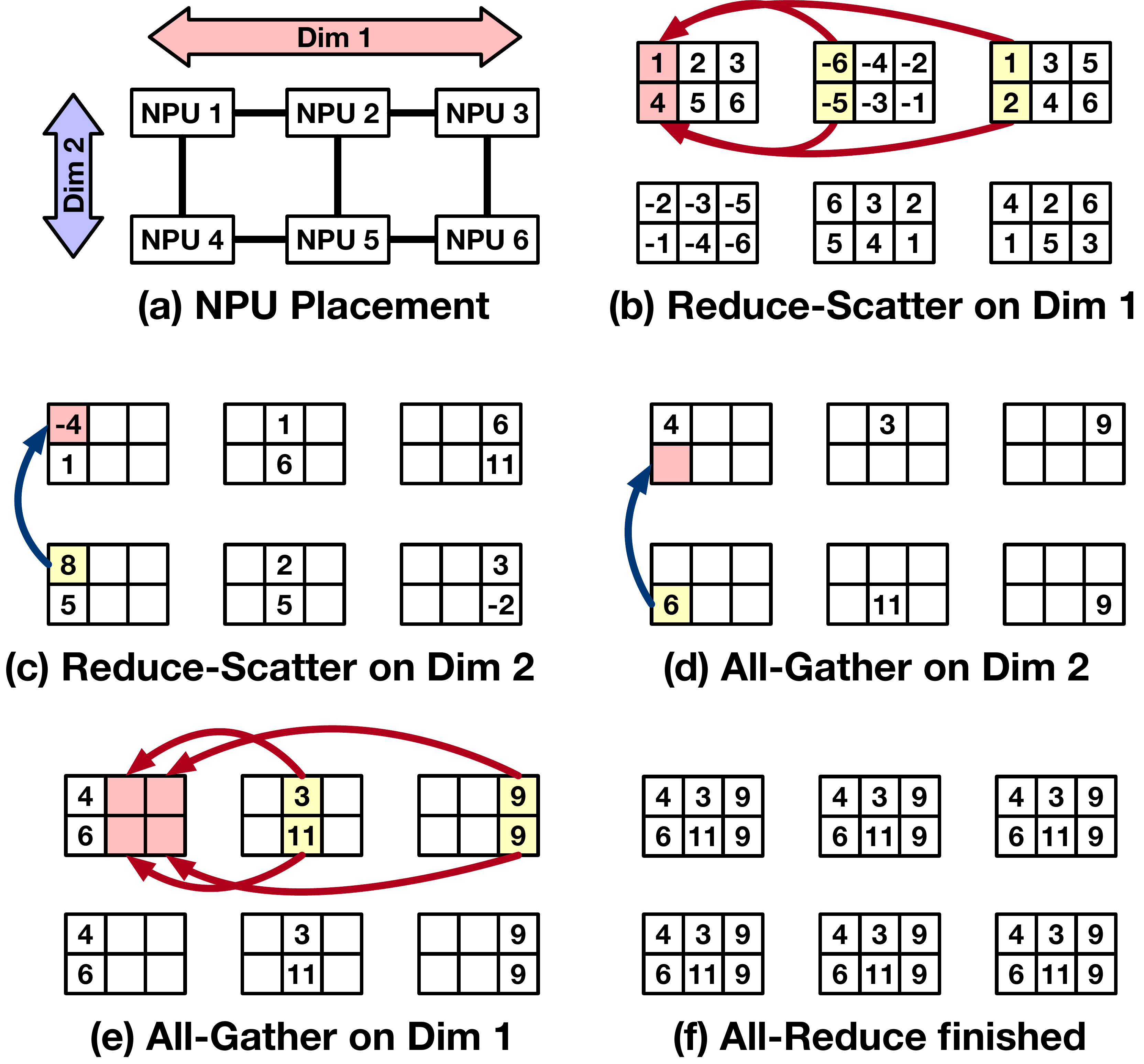}{
All-Reduce example on a $3\times2$ (2D) network. (a)~NPU placement (b--c)~Reduce-Scatter phase (d--e)~All-Gather phase (f)~All-Reduce collective finished. Arrows represent the traffic of chunks received by NPU 1 for each stage.
}{0.93}{-0.8em}{-1.5em}

\section{Motivation and Challenges}\label{sec:problemStatement}

\subsection{Technology Constraints of a Network Dimension}

ML Training platforms commonly utilize 2D networks that leverage two technologies: (i)~a high-BW proprietary network to connect multiple NPUs within a server node (i.e., \textit{scale-up}) and (ii)~NICs to connect NPUs across server nodes (i.e., \textit{scale-out})~\cite{dgx1,dgx2,NVIDIASuperPod,zionex}. However, increasing the raw network BW of each dimension in such setups is often challenging and expensive. This limitation arises due to the fundamental physical constraints imposed by the link technologies and their associated costs, such as pin counts, manufacturing, thermals, power, and area considerations~\cite{maximumNPUBW}. For instance, the current NVLink technology~\cite{NVLink3} can provide up to 450 GB/s per NPU, and the state-of-the-art NIC can drive up to 50 GB/s~\cite{NIC400G}.

\subsection{Opportunity: Multi-dimensional Networks}\label{subsubsec:motivation_multidim}

In large model distributed training, a multi-dimensional topology composed of multiple network technologies can be an effective solution. There are two key benefits that such topologies provide in the context of DNN training.

\noindent \textbf{Higher Aggregated BW.} Simultaneously leveraging multiple network technologies, including chiplets~\cite{simba} and photonics~\cite{tpuv4, passage}, offer an opportunity to further increase aggregate BW per NPU by introducing additional network dimensions between NPUs either on a package~\cite{simba} or on a board~\cite{tpuv4}.

\noindent \textbf{Higher Perf-per-cost.} As chunks are successively \textit{reduced} throughout the stages of the multi-rail \reducescatter\ phase, the volume of communication progressively decreases at each network dimension. This reduction in communication size is depicted in \autoref{fig:MultiDimCollExample}. NPU 1 first receives 4 chunks from its neighbors through Dim 1. However, since these messages are reduced into a single chunk, the second \reducescatter\ stage through Dim 2 only requires receiving 1 chunk. As a result, having more network dimensions significantly reduces the message size once it reaches higher network dimensions. This enables organizing a multi-dimensional network in a way that cheaper network technologies (i.e., scale-up) drive higher network BW resources in lower network dimensions. Expensive network technologies (e.g., NICs or photonics) can establish higher network dimensions while providing limited BW resources, benefiting from the reduced communication burden enabled by multi-dimensionality. Even when allocating equal BW resources per NPU, this architectural design yields significant perf-per-cost improvements by substituting expensive technologies with more cost-effective resources.

\insertFigure{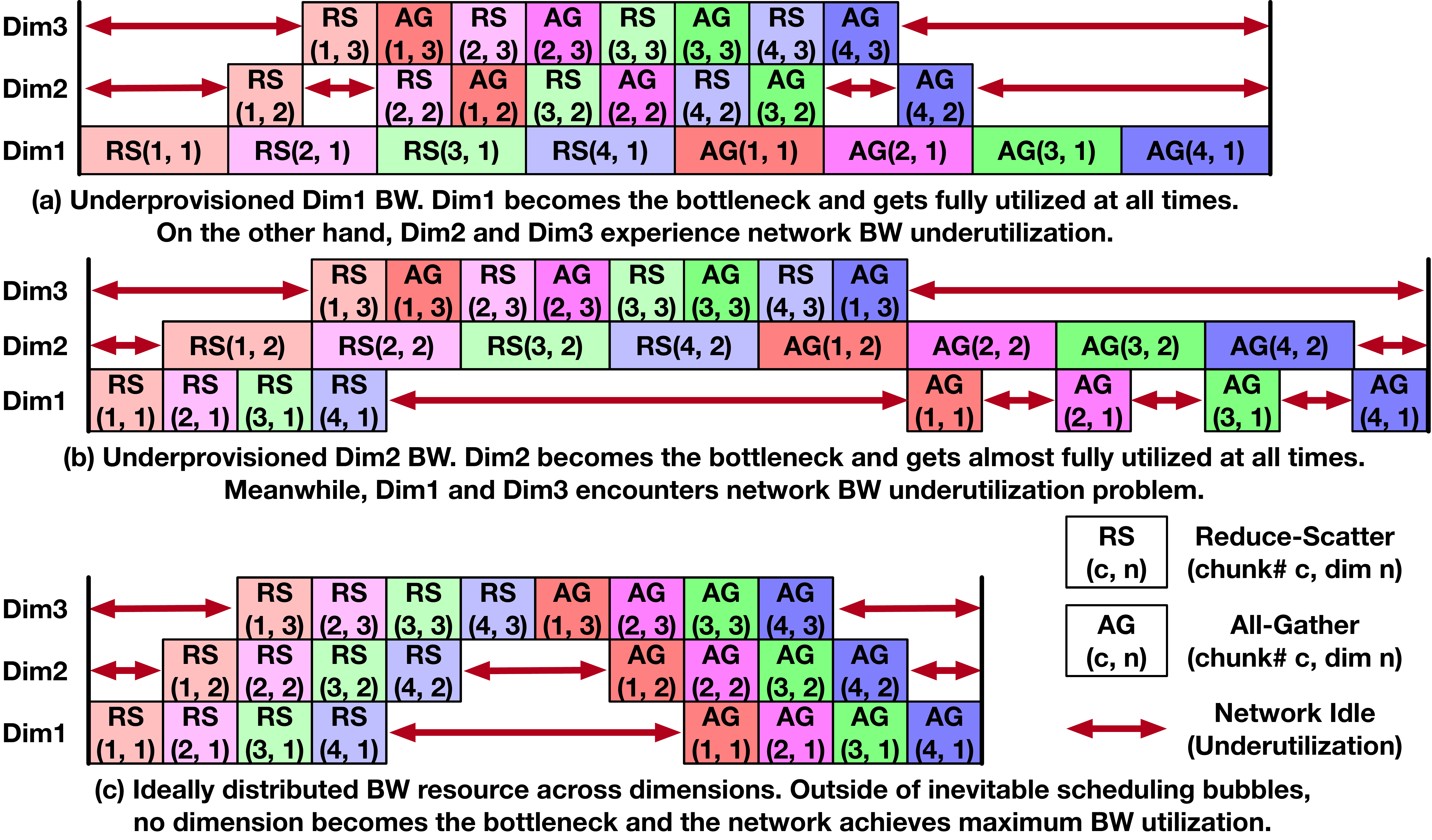}{
Running \allreduce\ with 4 chunks on 3D networks with different BW allocations.
}{1}{-2em}{-1.5em}

\subsection{Challenge: Design-time Consideration of Network BW}\label{subsubsec:problem_statement}

Realizing a multi-dimensional architecture is a challenging task since the allocation of physical BW across network dimensions at design time can significantly impact the runtime performance. Consequently, appropriately sizing the BW of each network dimension becomes crucial to avoid overprovisioning or underprovisioning network resources. This can be exemplified by the scenario in \autoref{fig:MultiDimCollExample}. The payload size of Dim 2 is only a quarter of Dim 1's due to the reduction phase in Dim 1. Consequently, the \textit{BW requirement} of Dim 2 is only 1/4 of Dim 1's. If the \textit{physical network BW} of Dim 2 is larger than 1/4 of Dim 1's, then Dim 2's BW resource is overprovisioned, leading to underutilization. Conversely, if Dim 2's BW is less than 1/4 of Dim 1's, Dim 2 becomes the bottleneck and Dim 1's network resource will be underexploited. A collective communication usually consists of multiple chunks that run in a pipelined manner~\cite{BLink}. \autoref{fig:BWAllocationMotivation} provides a generalized scenario by using a 3D network. In \autoref{fig:BWAllocationMotivation}(a), Dim 1's physical BW is smaller than its requirement, making Dim 1 the communication bottleneck and leading to significant underutilization of other dimensions (Dim 2 and Dim 3). Similarly, \autoref{fig:BWAllocationMotivation}(b) depicts a case where Dim 2's BW is underprovisioned, resulting in extensive network underutilization of Dim 1 and Dim 3. If we can judicially design a multi-dimensional network with such consideration, the network utilization can be maximized, as shown in \autoref{fig:BWAllocationMotivation}(c). The potential benefits of using workload-aware multi-dimensional networks are demonstrated in \autoref{fig:BaselineBWUtilization}. The plot shows the normalized runtime to train the \msft\ model, where each NPU has a 300 GB/s aggregated BW per NPU, but assuming different network BW utilization ratios. In the baseline \equalBW\ configuration (all network dimensions have an equal amount of BW, explained in \autoref{subsec:ExperimentalSetup}), the network BW utilization was only 57.53\%, 39.02\%, and 66.74\% for 2D, 3D, and 4D networks, respectively. By using a workload-aware network BW configuration optimized for the target \msft\ model, we can maximize the network BW utilization and theoretically achieve training speedups of 1.39$\times$, 1.83$\times$, and 1.29$\times$, respectively. This result emphasizes the critical role of network BW distribution across dimensions for AI collectives.

\insertFigure{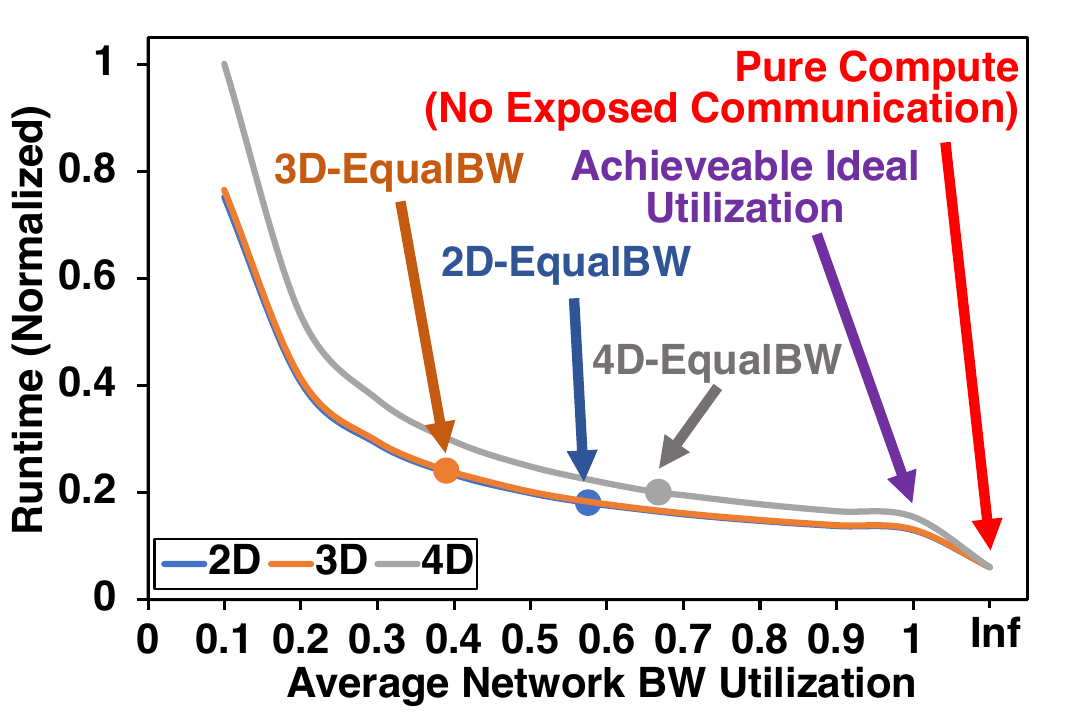}{
The normalized end-to-end training time of \msft\ on 2D, 3D, and 4D topologies with 300 GB/s per NPU. Average network BW utilization was 53.11\% (66.74\% max). We can achieve 1.83$\times$ maximum speedup if we can reach 100\% BW utilization.
}{0.72}{-1em}{-0.5em}

\insertFigure{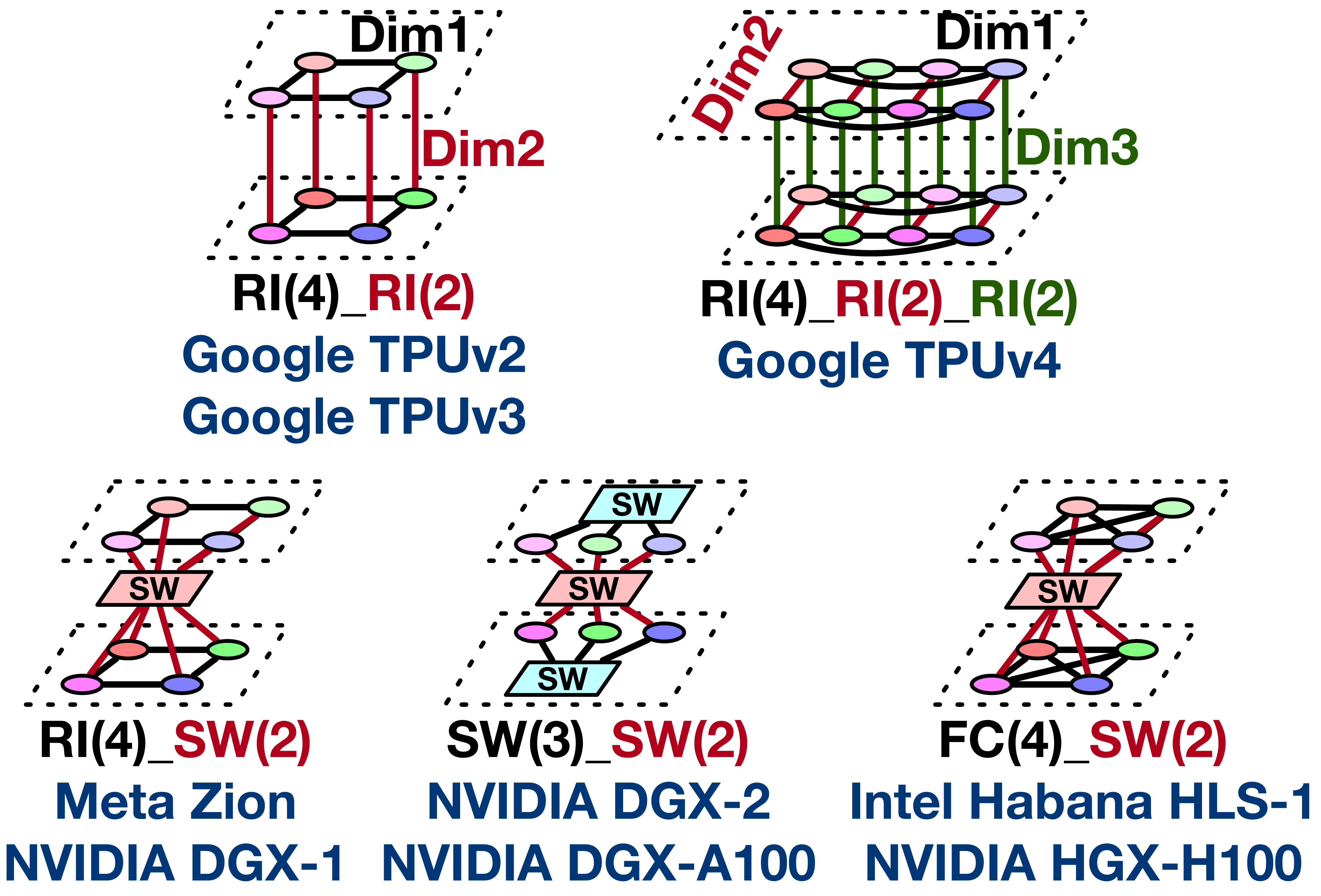}{
Examples of 2--3D networks and their corresponding names in the notation used in this work are provided. To demonstrate the expressiveness of the representation in capturing real systems, corresponding ML HPC clusters that utilize equivalent network shapes are also listed.
}{0.84}{-1em}{-1.5em}

\section{\framework}\label{sec:BWAllocation}

In this section, we introduce \framework: a workload-aware, design-time multi-dimensional network optimization framework. We explain how \framework\ can model collective communications, large model distributed training, and optimize the network BW of multi-dimensional topologies. \framework\ repository is open-sourced and can be publicly accessed.\footnote{
https://github.com/astra-sim/libra
}

\subsection{Multi-dimensional Network Representation}

In this work, we utilize the multi-dimensional network representation from \cite{astrasim2}. We adopt three unit topologies as building blocks for each network dimension: Ring~(RI), FullyConnected~(FC), and Switch~(SW), as shown in \autoref{fig:NetworkBuildingBlocks}. Multi-dimensional networks can be represented by stacking the building blocks together alongside the corresponding network size. For example, RI(4)\_RI(4)\_RI(4) represents a 3D Torus network with 64 NPUs. To demonstrate the capability to capture HPC networks in deployment, \autoref{fig:MultiDimTopologyExample} depicts five examples of 2--3D networks captured in this taxonomy and their corresponding ML HPC clusters in use. Although not shown, the notation is flexible and can be leveraged to capture multi-dimensional networks with four dimensions or more.
\subsection{Problem Statement and Constraints}

\framework\ is a design-time framework to construct an HPC network optimized for AI training. Given \textit{a set of target workloads} and a \textit{multi-dimensional network} representation, \framework\ performs optimization to yield the best network BW configuration. The optimization process aims to maximize specific objectives such as training performance or perf-per-cost. \framework, throughout optimization, ensures that the network BW configuration adheres to the user-defined design constraints, such as fixed BW per NPU, BW limit for a specific dimension, or limited total network cost.

\subsection{Modeling Distributed Training}\label{subsec:modeling_cc}

In order to optimize for the best BW configuration of a multi-dimensional network, \framework\ estimates the end-to-end training time as a function of network BW. To achieve this, \framework\ first models collective communications as a function of network BW. Consider a scenario where we run an $m$-sized \allreduce\ operation on a 2D ($n_1 \times n_2$) network with $B_1$ and $B_2$ for its two dimensions' BW. Mathematically, the nature of multi-rail \allreduce\ results in the traffic volume of $\frac{2m(n_1-1)}{n_1}$ and $\frac{2m(n_2-1)}{n_1 n_2}$ transferred by each dimension, respectively. Taking into account the BW of each dimension, and considering the bottlenecking dimension determines the collective time (as shown in \autoref{fig:BWAllocationMotivation}), the \allreduce\ time is $\max \big( \frac{2m(n_1-1)}{n_1 B_1}$, $\frac{2m(n_2-1)}{n_1 n_2 B_2}\big)$. Similarly, \reducescatter\ and \allgather\ can be modeled as $\max \big( \frac{m(n_1-1)}{n_1 B_1}$, $\frac{m(n_2-1)}{n_1 n_2 B_2}\big)$ since the communication volume halves compared to \allreduce. Finally, \alltoall\ time is computed as $\max \big( \frac{m(n_1-1)}{n_1 B_1}$, $\frac{m(n_2-1)}{n_2 B_2}\big)$ due to the absence of chunk reduction. Such models can be further generalized as the network dimensionality increases. Ultimately, \framework\ estimates all communication times as functions where the sole parameter is the network BW configuration.

The end-to-end model training can be captured by leveraging such communication representation. The training loop depicted in \autoref{fig:TrainingLoop}(b) has no overlap among the computation and communication stages. Therefore, the end-to-end execution time is estimated by simply aggregating all compute and communication delays. Speficially, the forward pass takes $\sum_{l \in \text{layer}} \big( \text{Fwd\_Compute}_l + \text{Fwd\_Comm}_l \big)$ time. Note that $\text{Fwd\_Compute}_l$ is BW-independent and $\text{Fwd\_Comm}_l$ is a function of network BW. Similarly, the backward pass takes $\sum_{l \in \text{layer}} \big( \text{TP\_Compute}_l + \text{TP\_Comm}_l + \text{DP\_Compute}_l + \text{DP\_Comm}_l \big)$. If there are overlaps in complex training loops, they can easily be reflected in the end-to-end time estimation. For example, training loop  \autoref{fig:TrainingLoop}(c) assumes that TP\_Compute is fully exposed but overlapping TP\_Comm with DP\_Compute and DP\_Comm is possible. Then, each layer's backward pass will take $\text{TP\_Compute}+\max(\text{TP\_Comm}, \text{DP\_Compute}+\text{DP\_Comm})$ time. The end result is \framework\ estimating and capturing the training performance of a DNN model as a function of network BW, which can then be optimized.

\noindent \textbf{\framework\ Modeling:} In the current \framework\ modeling, we disregarded the impact of link latencies or NPU performance implications (e.g., memory access BW or reduction performance). Large-model training commonly involves large traffic volumes, making communication highly network BW-bound~\cite{chen2023autoddl}. This allows modeling communication performance in network BW configuration to be a feasible option. In general, accurately modeling collective communications with more intricate equations has been the focus of recent endeavors, which is orthogonal to this work since \framework\ can incorporate such modeling and still optimize for network BW configurations.

\noindent \textbf{In-network Collective:} The offloading of collectives to the network switches has been a subject of research~\cite{CommBottleneck1} and commercial developments~\cite{NvidiaH100, MellanoxSHARP}. In essence, in-network collective offloading effectively reduces the communication time of Dim $i$ to $\frac{m}{n_1 n_2 ... n_{i-1} B_i}$.

\noindent \textbf{Parallelization Strategy:} Intricate parallelization strategies, including Pipeline Parallelism, periodically involve non-collective communication patterns such as direct NPU-to-NPU message transfers. Such operations could still be captured in terms of network BW (e.g., $\frac{m}{B_i}$) and can be leveraged in estimating end-to-end performance.

\begin{scriptsize}
\begin{table}[t]
\caption{Cost model for network cost evaluation. Dollar-cost of each network component per GB/s is shown. The values are derived from most current available data~\cite{NVIDIASuperPod,intelHabana,NIC400G,HabanaPtP, topoopt, terarack}. We used the lowest value of each entry for evaluation.}
\vspace{-1.6em}
\begin{center}
\begin{tabular}{|c|c|c|c|c|}
\hline
\textbf{(\$/GBps)} & \textbf{Link} & \textbf{Switch} & \textbf{NIC} \\ \hline
Inter-Chiplet & 2.0 & - & -  \\ \hline
Inter-Package & 4.0 -- 5.2 & 13.0 & -  \\ \hline
Inter-Node & 4.0 -- 5.2 & 13.0 & -  \\ \hline
Inter-Pod & 7.8 -- 16 & 18.0 -- 69.6 & 31.6 -- 144  \\ \hline
\end{tabular}
\label{table:unitCost}
\end{center}
\vspace{-1em}
\end{table}
\end{scriptsize}

\insertFigure{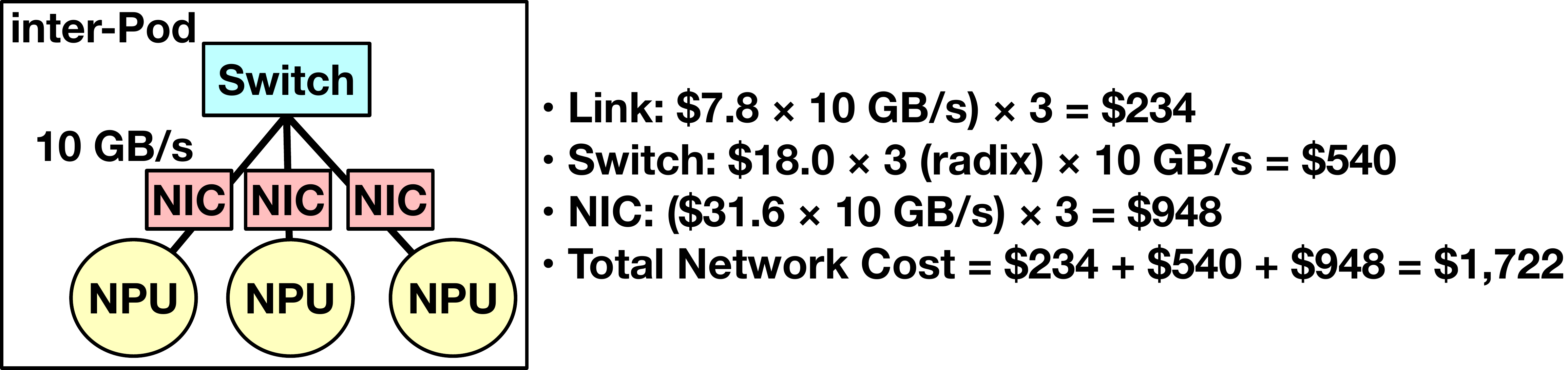}{
Example cost modeling of a 3 NPU switch network. The network Cost Model used for the cost modeling is shown in \autoref{table:unitCost}.
}{0.95}{-1em}{-1.5em}

\subsection{Modeling Network Cost}\label{subsec:model_network_cost}

In addition to performance modeling, \framework\ can estimate the dollar cost of multi-dimensional networks. This estimation process relies on the network cost model, which is provided as an input parameter to \framework\ by the user. The reason for this is that network costs can vary significantly based on the technology and specific vendors and may also change over time.\footnote{
For instance, the inter-node link costs can vary significantly across high-speed electrical~\cite{NVLink3, xelink} vs. photonic~\cite{tpuv4}.
} To facilitate the analysis, we offer a default cost model (shown in \autoref{table:unitCost}), derived from costs available in public citations. This default model is used for design-space exploration in this work. In this default cost model, Pod is defined as the unit for scale-out, meaning other network dimensions do not utilize NICs. For inter-Chiplet networks, we assume Chiplets are always connected peer-to-peer, thereby eliminating the need for switches. \autoref{fig:CostExample} depicts an example cost analysis based on this default cost model.

\subsection{Optimizing Multi-dimensional Network}

With the provided modeling capabilities, \framework\ can optimize the network BW for a target DNN training task, leading to the ideal network BW allocation that achieves specific objectives, such as maximizing training performance. \framework\ utilizes a Quadratic Programming (QP) solver~\cite{gurobi} to carry out the network optimization. Furthermore, \framework\ has the capability to optimize for a network that targets \textit{multiple target workloads}, accommodating real-world training scenarios.

\subsection{Optimization Scheme}\label{subsubsec:optimizationScheme}
\framework\ supports two optimization objectives to yield different network BW design points, as discussed below.

\noindent \textbf{(i) \perfBW.} The \perfBW\ optimization scheme aims to maximize the model training performance. \framework's QP solver disregards the network cost factor and the solver's optimization objective is set to minimize the end-to-end training time.

\noindent \textbf{(ii) \perfCostBW.} \perfCostBW\ takes into account both network cost and performance. It multiplies end-to-end training time by the estimated network cost, to measure the (reciprocal of) perf-per-cost metric, for which the QP solver inside \framework\ is then configured to minimize.

In practice, we envision that AI training clusters will be \textit{designed to handle a family of target workloads}, not just a single target. \framework\ is capable of designing a network aware of multiple target workloads. It is achieved by optimizing a \textit{weighted sum of the end-to-end training time} of individual target workloads, where the weight represents the importance of the corresponding workload. This results in a network design point that is optimized towards multiple workloads. By accommodating multiple targets and considering their relative importance, \framework\ becomes a versatile tool for designing networks that can effectively handle various training scenarios.

\framework\ harnesses the power of a QP solver to find an ideal network design point, providing the flexibility to represent linear and quadratic constraints as desired by the system designer. For instance, if there is a restriction on the total network BW per NPU, it can be easily represented as $\sum_{i} B_{i} \le B$, where $B_{i}$ denotes the BW of Dim $i$. Similarly, to limit the inter-Pod dimension BW to 50 GB/s, this constraint is captured as $B_{4} \le 50 \text{GB/s}$. The representation of constraints is highly adaptable, allowing for the addition of various flexible constraints of interest, such as $B_1 + B_2 = 500 \text{ GB/s}$ or $B_1 \ge B_2 \ge B_3$, $25 \le B_3 \le 150 \text{ GB/s}$. Multiple constraints can be applied simultaneously, enabling a comprehensive and customized network optimization.

\begin{scriptsize}
\begin{table}[t]
\caption{Workloads specifications used for analysis.}
\vspace{-1.6em}
\begin{center}
\begin{tabular}{|c|c|c|c|c|}
\hline
\textbf{Workload} & \textbf{Params} & \textbf{TP Size} \\ \hline
Turing-NLG & 17B & 1 \\ \hline
GPT-3 & 175B & 16 \\ \hline
MSFT-1T & 1T & 128 \\ \hline
DLRM & 57M (MLP layers only) & Across all NPUs \\ \hline
ResNet-50 & 25.6M & 1 \\ \hline
\end{tabular}
\label{table:workloads}
\end{center}
\vspace{-2em}
\end{table}
\end{scriptsize}

\section{Methodology}\label{sec:eval}

\subsection{Simulation Infrastructure}

\framework\ is a standalone framework aimed at aiding the design process for determining multi-dimensional network BW configurations. However, to profile and measure the performance of networks designed by \framework, a simulation infrastructure is necessary to meticulously identify the practical implications. In this paper, we utilized ASTRA-sim distributed ML simulator~\cite{astrasim, astrasim2} for the identification of the \framework -generated networks' training performance. ASTRA-sim enables executing complex DNN workloads, modeling complex network architecture, running various collective algorithms with chunk-level scheduling, and capturing compute-communication overlaps. 

\noindent \textbf{Validation.} ASTRA-sim is validated over multiple real ML systems and showed an error rate of 2.8--11.4\%~\cite{astrasimvalidation}.

\subsection{Experimental Setup}\label{subsec:ExperimentalSetup}

\begin{scriptsize}
\begin{table}[t]
\caption{
Multi-dimensional Topologies used for analysis.
}
\vspace{-1.6em}
\begin{center}
\begin{tabular}{|c|l|}
\hline
\textbf{Name} & \textbf{Shape} \\ \hline
4D-4K & RI(4)\_FC(8)\_RI(4)\_SW(32) \\ \hline
3D-4K & RI(16)\_FC(8)\_SW(32) \\ \hline
3D-512 & SW(16)\_SW(8)\_SW(4) \\ \hline
3D-1K & FC(8)\_RI(16)\_SW(8) \\ \hline
4D-2K & RI(4)\_SW(4)\_SW(8)\_SW(16) \\ \hline
3D-Torus & RI(4)\_RI(4)\_RI(4) \\ \hline
\end{tabular}
\label{table:target_topologies}
\end{center}
\vspace{-2.5em}
\end{table}
\end{scriptsize}

\noindent \textbf{Baseline.} In this work, we use EqualBW as a straw-person baseline. The EqualBW scheme distributes the given BW resource $B$ equally across all $N$ dimensions, resulting in $\frac{B}{N}$ allocated to each dimension. EqualBW is chosen as the baseline since it is a workload-agnostic allocation without performance considerations and an actual optimization process.

\noindent \textbf{Workloads.} We selected three transformer-based LLMs and two non-LLMs (recommendation and vision)~\cite{Transformer, ResNet, DLRM} for evaluations in this paper. To distribute these models, we assumed ZeRO phase-2 optimizers~\cite{mszero}. The specifications of the selected models are provided in \autoref{table:workloads}. All communications are split into 64 chunks per collective.

\noindent \textbf{Compute Model.} The average efficacy of the A100 GPU~\cite{A100} was measured to be 75\% (i.e., 234 TFLOPS) and was used to estimate NPU compute times in the evaluation.

\noindent \textbf{Network Topologies.} We adopted different target multi-dimensional networks from a related work~\cite{themis}, where we adjusted the last dimension size to scale the network size (ranging from 512 to 4,096 NPUs). The overall list of topologies evaluated in this work is summarized in \autoref{table:target_topologies}. Across case studies, as a representative configuration, we chose the 4D-4K topology. Additionally, for corresponding 3D topology evaluations, we utilized the 3D-4K network, which is created by combining two Ring dimensions of the 4D-4K topology.

\section{Result}\label{sec:result}
We conducted several comprehensive design space exploration case studies of multi-dimensional networks to showcase diverse use cases of \framework. Due to space constraints, we are highlighting the most interesting design points in each study.

\subsection{Sweeping Large Model Training}
Here, we utilize \framework\ to design a network optimized for a specific training workload.\footnote{
In practice, \framework\ is envisioned to design networks optimized for an ensemble of training models. This scenario is shown in \autoref{subsec:groupOpt}.
} 
We evaluated both 3D-4K and 4D-4K networks. Additionally, we swept over the BW per NPU in the range of 100--1,000 GB/s, as it can cover the BW budget of recent ML clusters~\cite{tpuv4, A100, NvidiaH100}.

\insertFigure{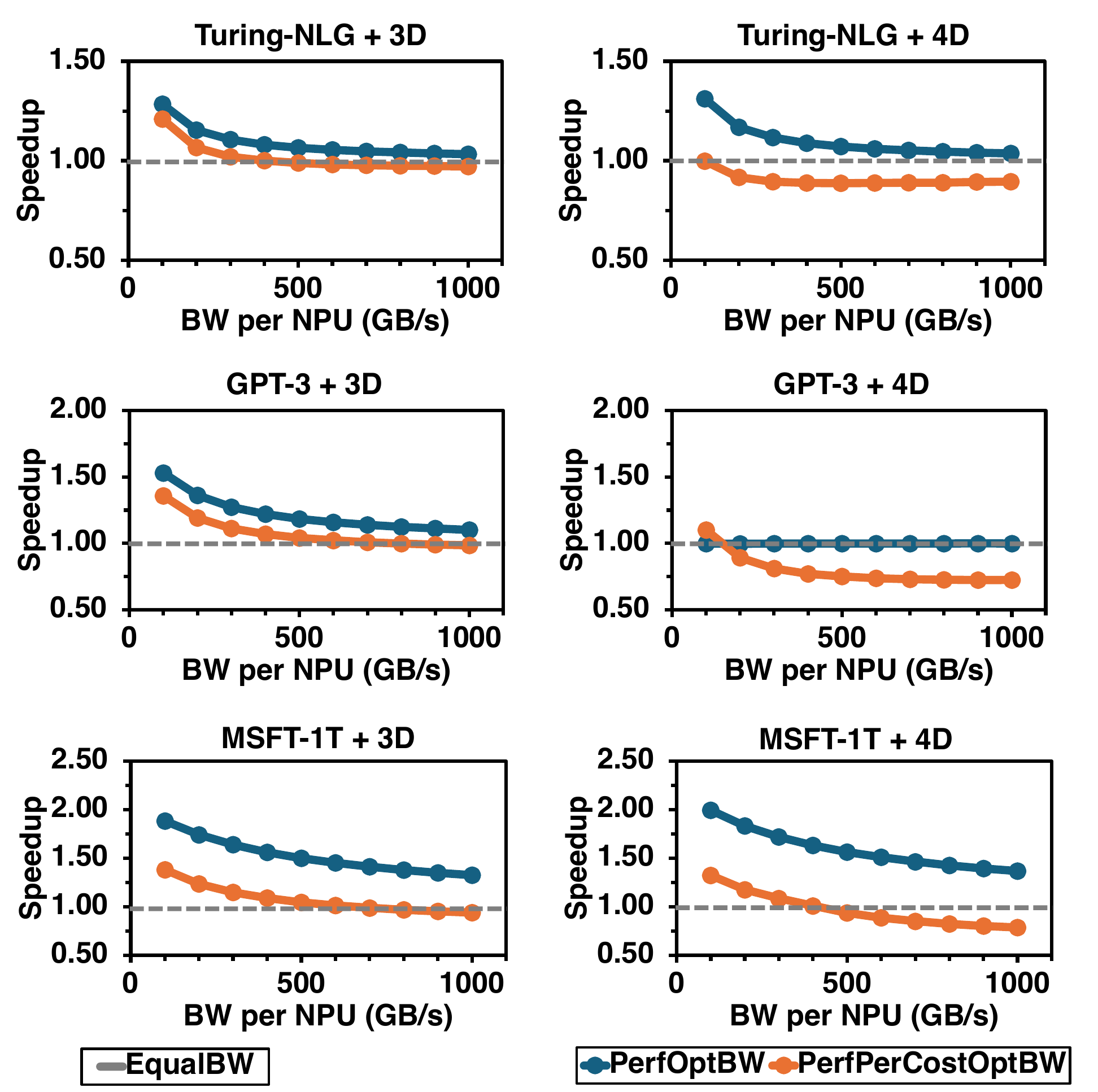}{
End-to-end training speedup over the baseline \equalBW. Each point represents a \framework-optimized network for the target workload. For example, \gpt+3D means a 3D network that is optimized for running \gpt. \perfBW\ network is optimized for maximizing training performance, while \perfCostBW\ maximizes perf-per-cost (thereby can have a speedup of less than 1).
}{1}{-2.3em}{-1.5em}

\noindent \textbf{Performance.} The performance benefit represented by the speedup over the baseline \equalBW\ configuration, for training \nlg, \gpt\ and \msft\ models over \perfBW\ and \perfCostBW-optimized networks, is depicted in \autoref{fig:PerfSweep}. Comparing the BW optimization schemes, we observe that \textit{\perfBW\ consistently provides the best performance}. On average, \perfBW\ achieves a 1.23$\times$ speedup (2.00$\times$ max) compared to the \equalBW\ network. Note that for \gpt\ on the 4D-4K topology, the training process cannot leverage all Dim 2 BW resources \framework\ assigned, due to the mismatching TP size, thereby yielding performance close to the baseline. Nonetheless, \perfBW\ for this configuration still achieved a 4.58$\times$ perf-per-cost improvement compared to the baseline.

\noindent \textbf{Perf-per-cost.} The perf-per-cost measurement relative to the baseline \equalBW\ network, is demonstrated in \autoref{fig:PerfPerCostSweep}. \perfCostBW\ achieves the highest perf-per-cost for all design points. On average, the perf-per-cost benefit of \perfBW\ and \perfCostBW\ over the baseline is 5.40$\times$ (12.24$\times$ max) and 9.16$\times$ (13.02$\times$ max), respectively.

\insertFigure{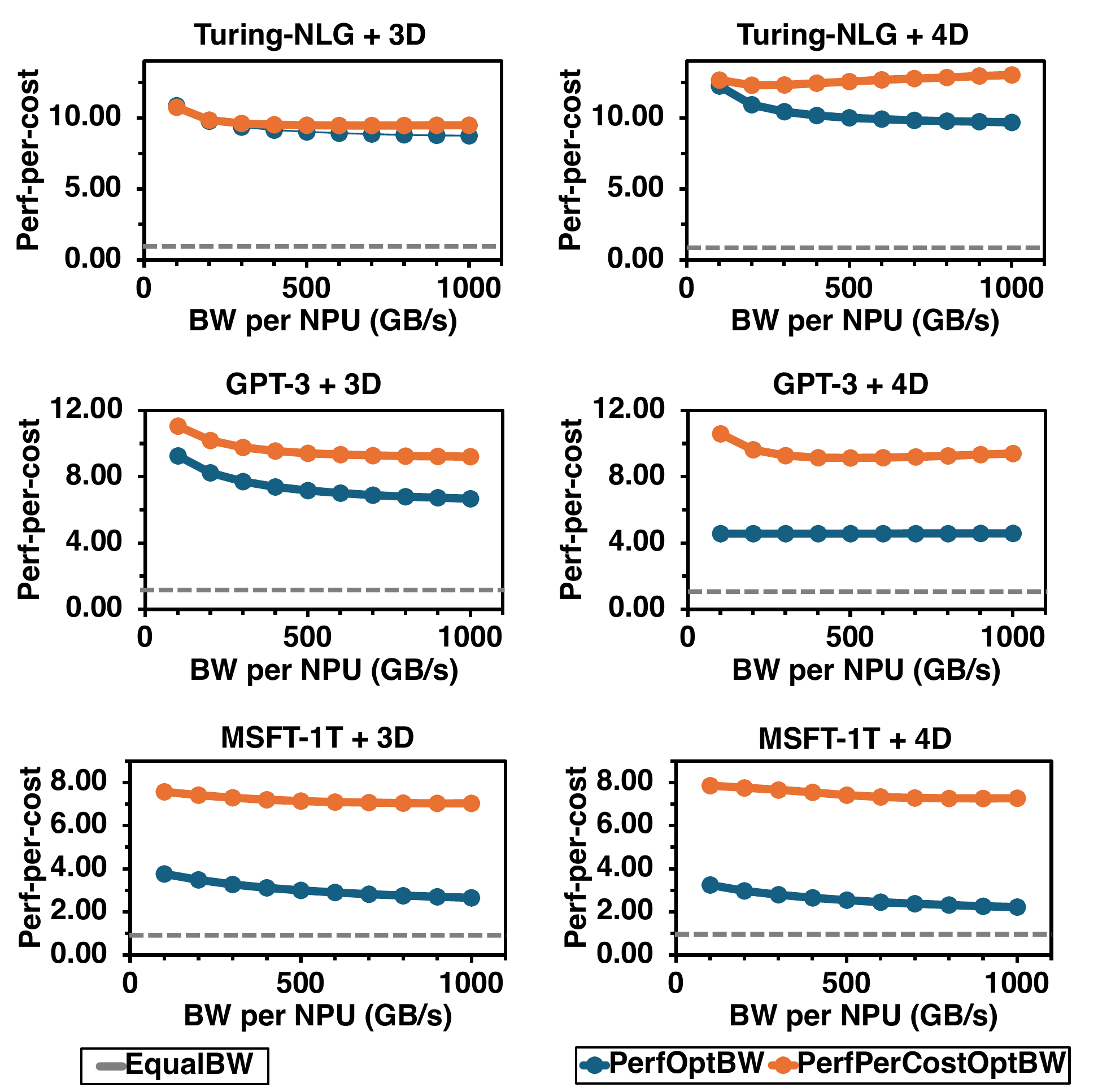}{
Perf-per-cost benefit of \framework\ over the \equalBW\ baseline.
}{1}{-2.2em}{-0.5em}

\insertFigure{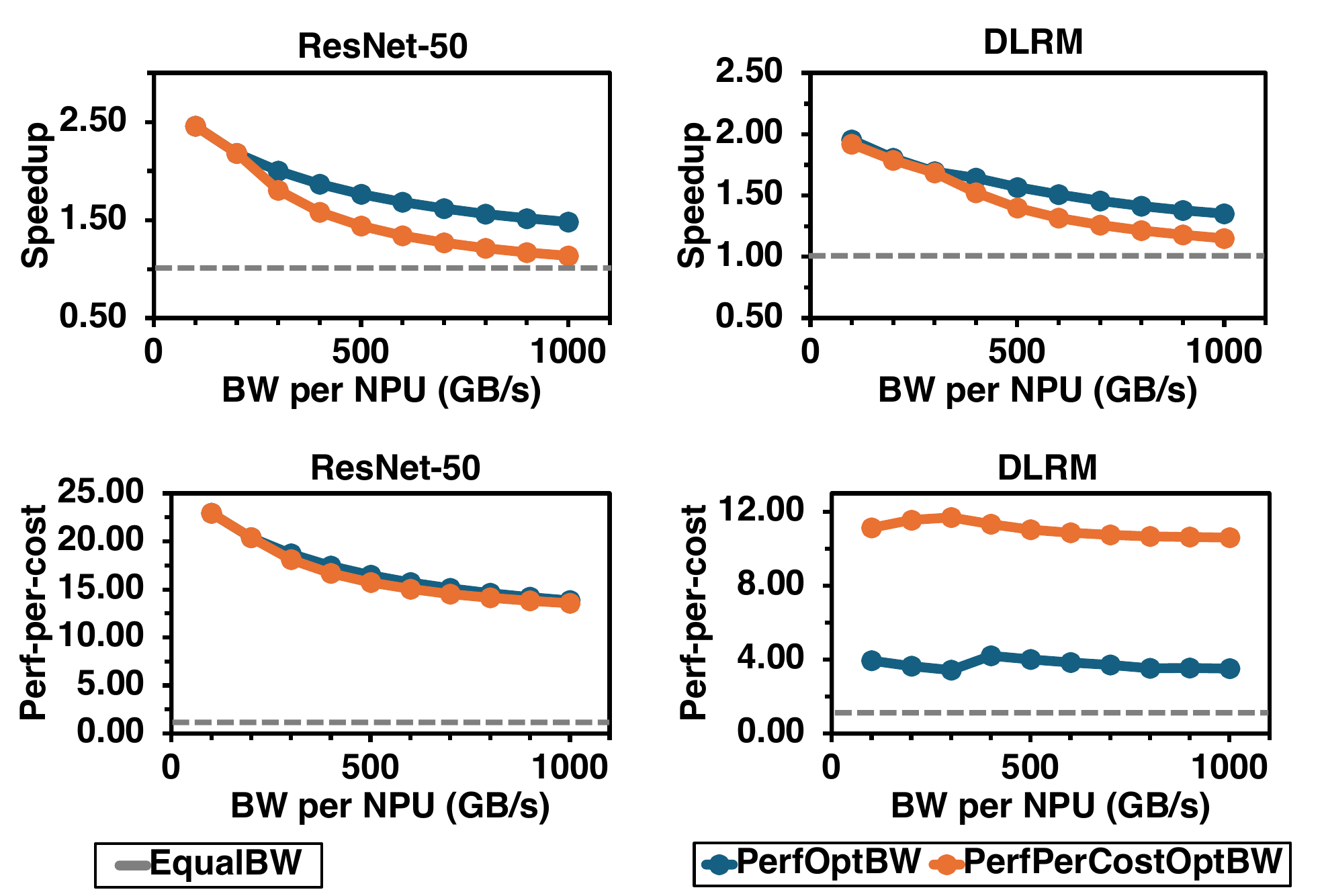}{
Speedup and perf-per-cost analysis of \resnet\ and \dlrm\ model. Results are normalized over corresponding \equalBW\ baseline.
}{1}{-2.2em}{-1.5em}

\noindent \textbf{Key Insights.} Larger models exhibit more performance benefits while smaller workloads show higher perf-per-cost advantages. Smaller models are less communication-critical, therefore, optimizing the network has a limited impact on performance. As workload sizes tend to increase, judicious design time considerations of \framework\ would gain more importance.

\noindent \textbf{Non-Transformer Workloads.} \framework\ can optimize the network for non-transformer workloads without any modification. For \resnet\ and \dlrm, we optimized the 4D-4K network using both \perfBW\ and \perfCostBW\  objectives. The results are summarized in \autoref{fig:ResultNonTransformer2x2}. It is worth noting that the \resnet\ model, due to its relatively small size compared to large LLMs, yields very small numbers as its estimated training time. This makes the perf-per-cost metric heavily cost-dependent. Due to the intrinsic numerical instability of the QP solver, \perfCostBW\  generated network designs with similar perf-per-cost to \perfBW. Nonetheless, \perfCostBW\  takes the network cost into consideration during the optimization process, resulting in network designs that are 15.41\% cheaper on average compared to \perfBW.

\noindent \textbf{Topology Exploration}. \framework\ supports networks of various shapes and scales. In order to showcase how \framework\ operates for a variety of target networks, we also compared three distinct networks with different dimensionality, size, and shape implications: 3D-512, 3D-1K, and 4D-2K. We ran \framework\ using the \msft\ workload and compared the speedup and perf-per-cost benefit over their corresponding \equalBW\ baseline. The results are summarized in \autoref{fig:TopologyCaseStudy}.

\insertFigure{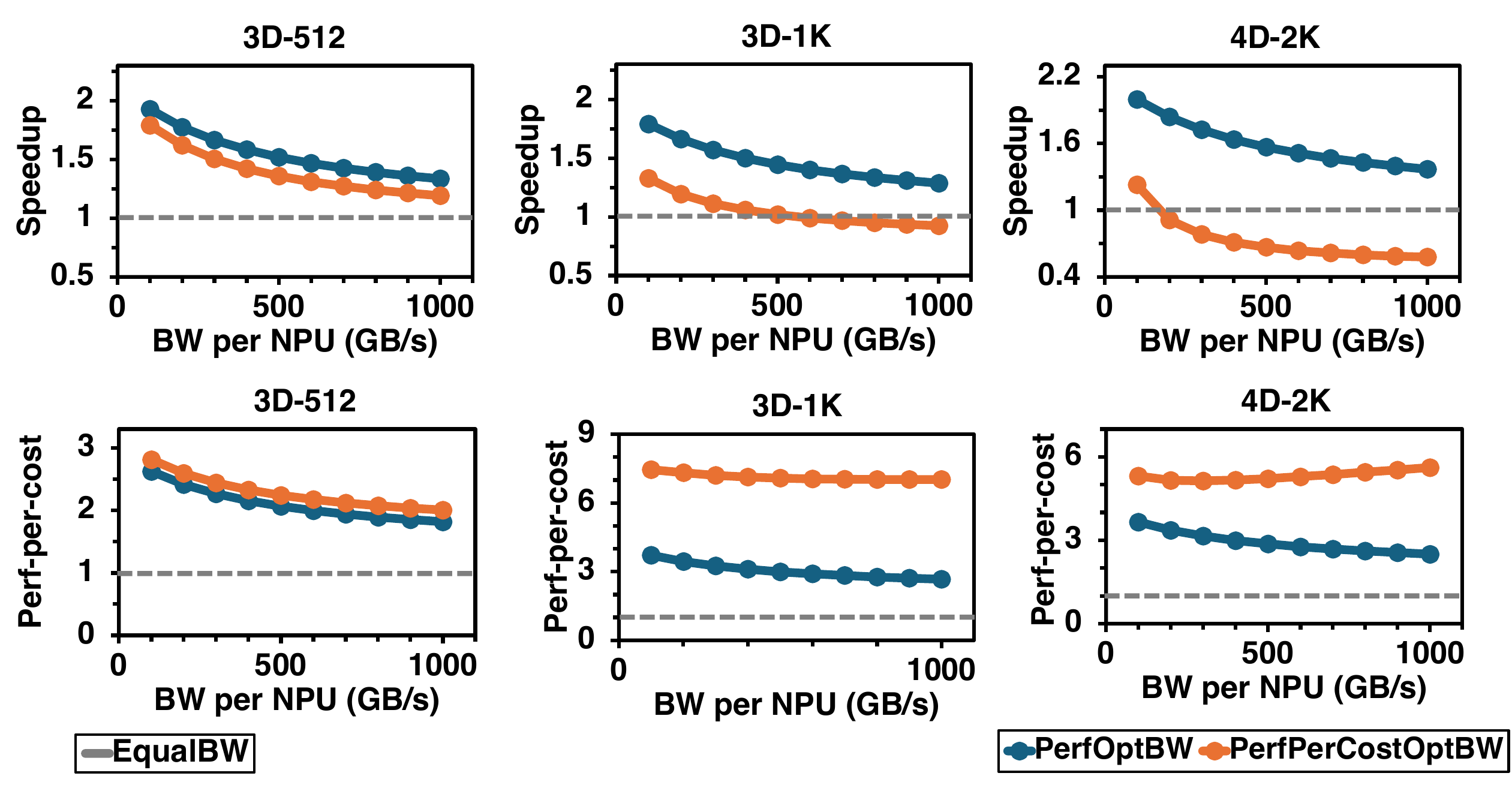}{
Speedup and perf-per-cost analysis of \msft\ over 3D-512, 3D-1K, and 4D-2K topologies. Results are normalized over their corresponding \equalBW\ baseline.
}{1}{-2.2em}{-0.5em}

\insertFigure{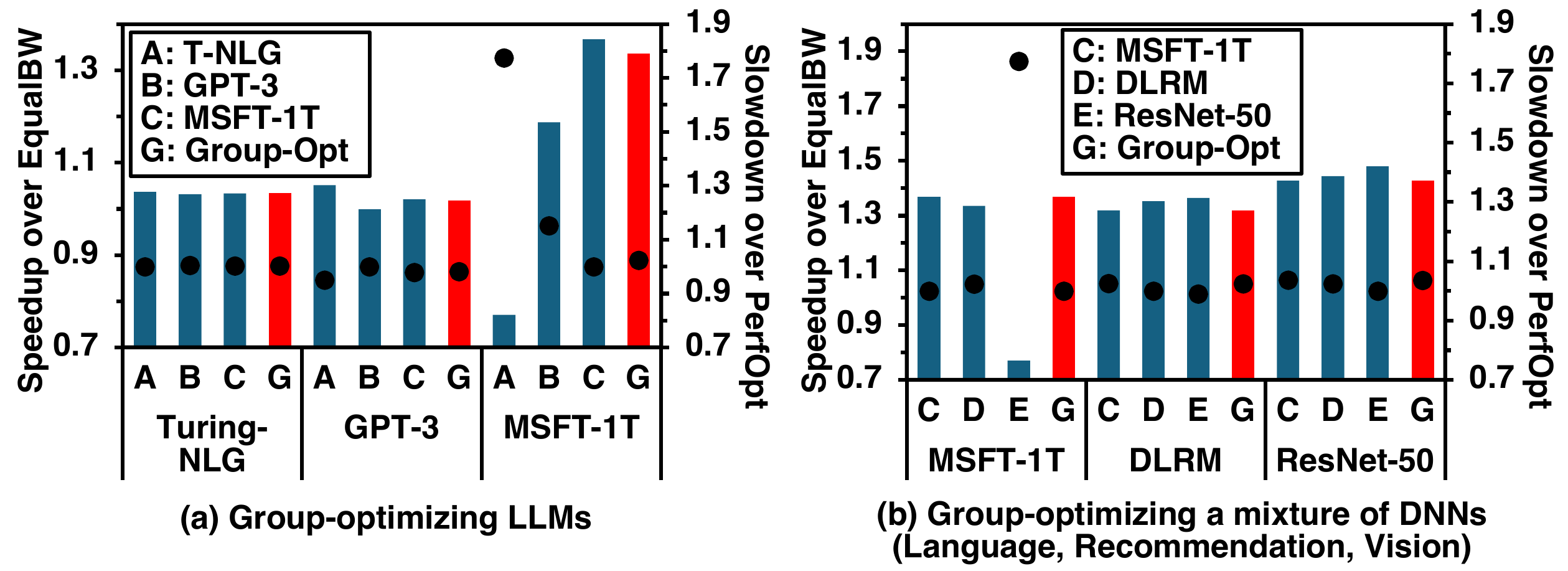}{
Speedup (over \equalBW, in bars) and slowdown (over \framework-optimized network, in dots) of various DNN models. The red bars denote performance over the group-optimized network. (a)~group-optimization among LLMs (b)~group-optimization across a mixture of DNNs.
}{1}{-2.2em}{-1.5em}

\subsection{Optimizing for Multiple Workloads}\label{subsec:groupOpt}
\framework\ can assist in designing a network by considering a group of workloads, rather than just a single target. For the 4D-4K network with 1,000 GB/s per NPU, we ran \perfBW\ to design networks optimized for specific targets. Then, we ran other non-targeted workloads on these networks to observe slowdowns compared to the training time over the optimized network. Additionally, we evaluated the network designed by \framework\ by considering all target networks at once during the optimization (i.e., group-optimization). We summarize the speedup (over baseline \equalBW) and slowdown (over \framework-optimized networks) in \autoref{fig:GroupOptFinal}. When we optimized the network separately for \resnet, \dlrm, \nlg, \gpt, and \msft, and performed non-optimized model training over them, we observed slowdowns of up to 1.77$\times$, 1.02$\times$, 1.77$\times$, 1.15$\times$, and 1.04$\times$, respectively. However, when leveraging all target models jointly to optimize the network, this network achieved near-optimal performance across all workloads, achieving the average slowdown of only 1.01$\times$.

\subsection{Cost Model Sensitivity Analysis}

\framework's cost model is a user-defined input, offering maximum flexibility since the cost model can change as technology evolves. To showcase this flexibility, we conduct a sensitivity analysis by varying the inter-Package link costs. We evaluated 4D-4K network with a BW budget of 1,000 GB/s per NPU. We used \perfCostBW\ while changing the inter-Package link cost from \$1--\$5/GBps. The resulting perf-per-cost (over baseline \equalBW) is depicted in \autoref{fig:CostSweepResult}. On average, the benefit over the \equalBW\ is $4.06\times$ ($5.59\times$ max).

\subsection{\framework\ with Runtime Optimizations}

Since the benefits of runtime-based optimizations are fundamentally determined by their underlying system configurations, incorporating careful design time considerations using \framework\ can enhance their efficacy.

\noindent \textbf{\framework+Themis.} Themis~\cite{themis} is a recently proposed collective scheduler that optimizes network BW utilization. It dynamically schedules data chunks over multi-dimensional networks in a greedy-based manner. To measure the joint benefit, we trained \gpt\ with Themis using 4D-4K topology, one configured with \equalBW\ and the other using \framework. We evaluated two distinct setups:

\begin{itemize}[leftmargin=*]
    \item iso-cost: both topologies have the cost of \$15M
    \item iso-resource: both topologies have 1,000 GB/s per NPU
\end{itemize}

The normalized performance and perf-per-cost are shown in \autoref{fig:ResultLibraThemis}. For iso-cost, the \framework-designed network was able to support 5.05$\times$ more BW per NPU compared to the \equalBW\ network. Consequently, even with Themis enabled for both systems, the \framework-designed network yielded a 2.24$\times$ training speedup. For iso-resource, the \framework-designed network showed a 1.04$\times$ better performance over the \equalBW-configured topology, while achieving 4.58$\times$ network cost reduction, leading to a 4.77$\times$ better perf-per-cost.

\insertFigure{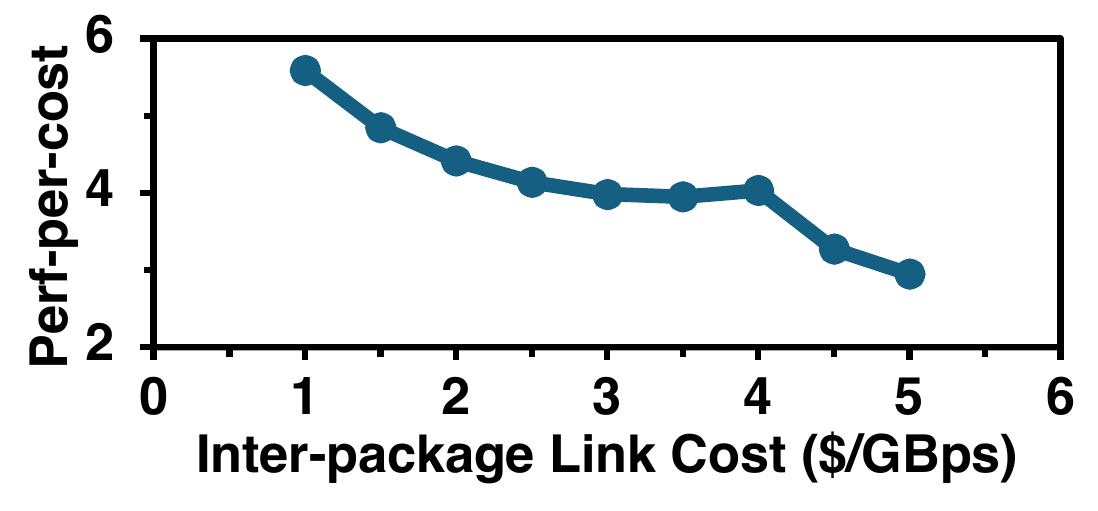}{
Normalized perf-per-cost (over baseline \equalBW) benefit of \perfCostBW\ on the 4D-4K network, while sweeping the inter-Package link cost from \$1/GBps to \$5/GBps.
}{0.7}{-1.2em}{-0.5em}

\insertFigure{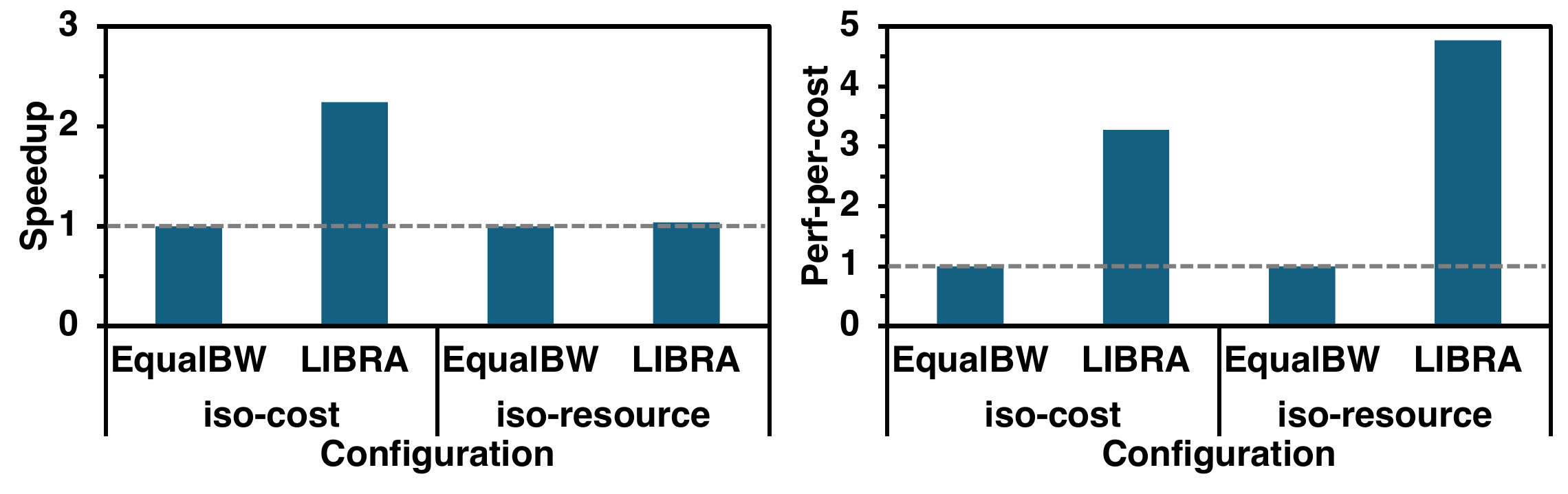}{
Normalized performance and perf-per-cost of \framework-optimized 4D-4K network (over \equalBW), when Themis runtime optimizer is leveraged for both networks.
}{1}{-1em}{-1.5em}

\noindent \textbf{\framework+TACOS}.
TACOS~\cite{won2023tacos} is a runtime-based tool that can automatically synthesize topology-aware collective algorithms. To demonstrate the co-design opportunities, a 1GB \allreduce\ with 8 chunks was synthesized and run over the 3D-Torus network at 1,000 GB/s per NPU. The normalized performance and perf-per-cost are shown in \autoref{fig:ResultLibraTACOS}. \framework+TACOS showed 1.25$\times$ and 1.08$\times$ speedup over the \framework-only and TACOS-only, respectively. Notably, due to the dollar-cost benefit from \framework, \framework+TACOS exhibited a 1.36$\times$ better perf-per-cost over the TACOS-only setup.

\subsection{Co-optimizing Network and Parallelization}

\framework\ offers the opportunity for joint co-optimization of the network and target workloads. In this study, we explore the impact of co-optimizing both the parallelization strategy and the network. To investigate this, we modify the TP-DP sizes, resulting in various parallelization strategies for the 4D-4K network with 1,000 GB/s per NPU. For this experiment, we relax the constraint of limited NPU memory and instead assume that the NPUs have access to an extended memory capacity through technologies such as CXL~\cite{CXL} or CPU memory~\cite{DeepSpeed}. In \autoref{fig:ResultCoOptimization}, we present the performance of \framework's \perfBW\ network normalized over the baseline using the \equalBW\ network with HP-(128, 32) parallelization, while varying the parallelization strategies between HP-(8, 512) and HP-(256, 16) for the \msft\ workload. As depicted in the figure, we achieve peak training performance with the 4D-4K network using HP-(64, 64). For the co-optimized 4D-4K network, \msft\ training was 1.19$\times$ faster compared to the baseline configuration.

\noindent \textbf{Key Insights.} Training performance significantly degrades when the TP size decreases below 32. This outcome arises from the intricate interplay between TP and DP communication for different parallelization strategies. The total network communication size is minimized when using HP-(32, 128). However, for HP-(64, 64), \framework\ was still able to configure a network that significantly reduced DP time while mitigating the increase in TP time. For other parallelizations, the total communication size was significantly higher due to increased TP or DP communications, resulting in reduced training performance even with \framework\ support.

\insertFigure{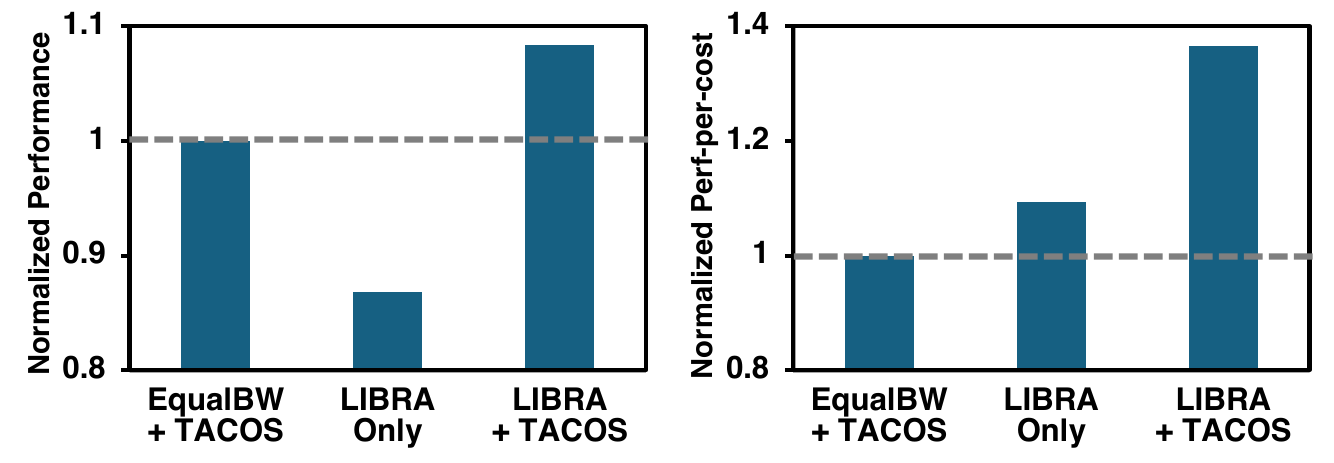}{
Normalized performance and perf-per-cost when jointly leveraging TACOS runtime collective synthesizer. \framework-only and \framework+TACOS results are normalized over \equalBW+TACOS.
}{1}{-2.3em}{-1.5em}

\insertFigure{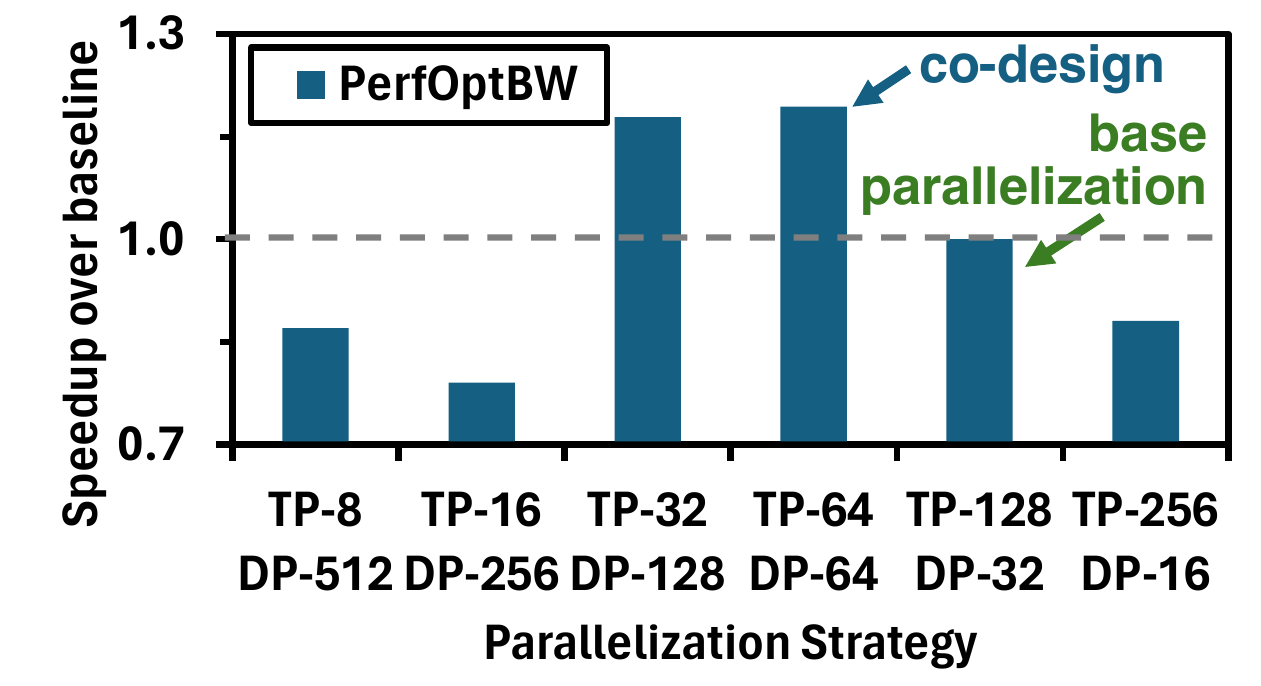}{
Normalized speedup of \msft\ across various parallelization strategies. All results are normalized to \equalBW\ with HP-(128, 32). HP-(64, 64) parallelization, along with its co-optimized \perfBW\ network, demonstrates the best training performance.
}{0.8}{-1.2em}{-1.5em}

\section{Related Work}\label{sec:related}

\noindent \textbf{Multi-dimensional Networks.} Multi-dimensional networks have been utilized in general-purpose HPC platforms, including 3--6D networks~\cite{ibm2, ibm1, tofu}. However, these networks were not specifically designed for AI training. In the AI domain, 3D~\cite{A100, NvidiaH100, tenstorrent} and 4D~\cite{googleCloudTpu} networks are being explored.

\noindent \textbf{Design-Time Optimization.} Collectives differ from point-to-point messages as all NPUs communicate synchronously, and message sizes change throughout the process. Such communication behavior becomes even more complex when using multi-dimensional algorithms. Various topology-aware collective algorithms have been proposed~\cite{mpi1, mpi2, mpi3, mpi4, mpi5, mpi6}; however, they are not specifically AI-aware and do not consider workloads, parallelizations, training loops, or network constraints. To the best of our knowledge, this is the first work targeting multi-dimensional network BW optimization for AI workloads at design-time. Another set of proposals offloads collectives to network switches~\cite{CommBottleneck1,MellanoxSHARP} or NICs~\cite{saeedACE}, which are orthogonal to \framework. We have incorporated these offload features in our modeling to find the best network BW.

\noindent \textbf{Co-optimization.} AI HPC clusters are being co-optimized alongside their target workloads due to the substantial resource demand. EFLOPS~\cite{EFLOPS} proposes both a novel network topology and collective algorithm to optimize \allreduce\ collectives. ZionEx~\cite{zionex} is a training cluster specifically aimed at accelerating DLRM training tasks. We also demonstrate that runtime-based optimizations perform at their best when leveraged with careful design-time considerations using \framework.

\section{Conclusion}

In this paper, we propose \framework, a design-time framework to construct workload-aware multi-dimensional networks. To the best of our knowledge, this is the first framework in the field of ML to optimize the BW of a multi-dimensional network, at design-time, using target workload characteristics. \framework\ aims to maximize the training performance and perf-per-cost of a target set of DNN workloads.

\section*{Acknowledgment}

This work was supported through awards from Intel. Additionally, this work has been developed and maintained with the support from Semiconductor Research Corporation~(SRC), the SRC AIHW program, as well as ACE Center for Evolvable Computing, one of the seven centers of the SRC JUMP 2.0 program. We also extend our sincere appreciation to the reviewers for dedicating their time and providing insightful comments that contributed to the improvement of this paper.

\bibliographystyle{IEEEtran}
\bibliography{reference/reference.bib}

\end{document}